\documentclass[10pt,twocolumn]{article}
\usepackage[top=1.9cm, bottom=1.9cm, left=1.8cm, right=1.8cm]{geometry}

\usepackage{amsmath,amssymb,amsfonts}
\usepackage{listings}                       %
\usepackage[noend]{algpseudocode}
\usepackage{multirow}
\usepackage{graphicx}
\usepackage{textcomp}
\usepackage{xcolor}
\usepackage{flushend}
\usepackage[lined,boxed]{algorithm2e}
\usepackage[bf]{caption}
\usepackage{graphicx,subcaption}

\usepackage[square,sort,comma,numbers]{natbib}

\SetCommentSty{mycommfont}

\usepackage{xcolor}
\usepackage{booktabs}
\usepackage{graphicx}
\usepackage{paralist}

\usepackage{times}

\usepackage[compact]{titlesec}
\titlespacing*{\section}{0pt}{*4}{4pt}
\titlespacing{\subsection}{0pt}{*3}{3pt}
\titlespacing{\subsubsection}{0pt}{*3}{3pt}

\definecolor{linkcol}{rgb}{0,0,0.5}
\definecolor{citecol}{rgb}{0,0.5,0.3}
\definecolor{urlcol}{rgb}{0.3,0,0}

\usepackage{xspace}

\renewenvironment{thebibliography}[1]{
  \begin{oldthebibliography}{#1}
    \setlength{\itemsep}{0.0em}
    \setlength{\parskip}{0.0em}
}
{
  \end{oldthebibliography}
}

\usepackage[hang,flushmargin]{footmisc}

\renewcommand{\footnoterule}{%
  \kern -3pt
  \hrule width 1in
  \kern 2pt
}

\usepackage{url}
\makeatletter
\def\url@leostyle{%
  \@ifundefined{selectfont}{\def\UrlFont{}}%
  {\def\UrlFont{}}%
}
\makeatother
\usepackage[hyphenbreaks]{breakurl}

\definecolor{darkred}{RGB}{153,0,0}
\definecolor{darkblue}{RGB}{0,0,99}
\usepackage[colorlinks=true, linkcolor = darkred,   citecolor = darkred, urlcolor = darkblue]{hyperref}

\newtheorem{ldp-definition}{Definition}
\newtheorem{dp-definition}{Definition}
\usepackage{enumitem}

\usepackage[framemethod=TikZ]{mdframed}

\mdfdefinestyle{MyFrame}{
    linecolor=black,
    outerlinewidth=0.5pt,
    roundcorner=5pt,
    innertopmargin=2pt,
    innerbottommargin=2pt,
    innerrightmargin=-10pt,
    innerleftmargin=-15pt,
    leftmargin = -10pt,
    rightmargin = -10pt,
    backgroundcolor=gray!10,
    userdefinedwidth=220pt
}

\newif
\ifcomment
\commenttrue
\ifcomment
\newcommand{\sz}[1]{{\bf \textcolor{brown}{SZ: #1}}}
\newcommand{\edc}[1]{{\bf \textcolor{green}{EDC: #1}}}
\newcommand{\gs}[1]{{\bf \textcolor{red}{GS: #1}}}
\newcommand{\ppaudel}[1]{{\bf \textcolor{blue}{PP: #1}}}
\newcommand{\jbnote}[1]{{\bf \textcolor{magenta}{JB: #1}}}
\else
\newcommand{\sz}[1]{}
\newcommand{\edc}[1]{}
\newcommand{\gs}[1]{}
\newcommand{\ppaudel}[1]{}
\newcommand{\jbnote}[1]{}
\fi

\newcommand{\systemname}{\textsc{Lambretta}\xspace}
\newcommand{\totalclaims}[1]{300\xspace}
\newcommand{\totalcandidatetweets}[1]{100,000\xspace}
\newcommand{\basekeyword}[1]{\textbf{\textit{base keywords}}}
\newcommand{\spanningsubset}[1]{\textbf{\textit{spanning subset}}}
\newcommand{\initialtrainmodel}[1]{\textbf{\textit{initial train model}}}
\newcommand{\initialtrainclaims}[1]{\textbf{\textit{archival train claims}}}
\newcommand{\validationclaims}[1]{\textbf{\textit{archival validation claims}}}
\newcommand{\twittervalidationclaims}[1]{\textbf{\textit{VoterFraud validation claims}}}
\newcommand{\expandedclaims}[1]{\textbf{\textit{expanded claims}}}
\newcommand{\expandedmodel}[1]{\textbf{\textit{expanded model}}}
\newcommand{\descr}[1]{\smallskip\noindent\textbf{#1}}
\newcommand{\descrit}[1]{\vspace{0.05cm}\noindent\emph{#1}}
\newcommand{\done}[1]{$\textbf{G}_1$\xspace}
\newcommand{\dtwo}[1]{$\textbf{G}_1$\xspace}
\newcommand{\dthree}[1]{$\textbf{D}_1$\xspace}
\newcommand{\dfour}[1]{$\textbf{D}_2$\xspace}

\def\BibTeX{{\rm B\kern-.05em{\sc i\kern-.025em b}\kern-.08em
    T\kern-.1667em\lower.7ex\hbox{E}\kern-.125emX}}

\begin{document}

\sloppy

\title{\bf \systemname: Learning to Rank for Twitter Soft Moderation\thanks{To appear in the Proceedings of the 44th IEEE Symposium on Security and Privacy (S\&P 2023). Please cite accordingly.}}

\author{Pujan Paudel$^{1}$, Jeremy Blackburn$^{2}$, Emiliano De Cristofaro$^{3}$, Savvas Zannettou$^{4}$, and Gianluca Stringhini$^{1}$\\[1ex]
\normalsize $^{1}$Boston University, $^{2}$Binghamton University, $^{3}$University College London, $^{4}$Delft University of Technology\\
\normalsize \{ppaudel,gian\}@bu.edu, jblackbu@binghamton.edu, e.decristofaro@ucl.ac.uk, s.zannettou@tudelft.nl\vspace*{-0.3cm}}\date{}

\maketitle

\begin{abstract}

To curb the problem of false information, social media platforms like Twitter started adding warning labels to content discussing debunked narratives, with the goal of providing more context to their audiences.
Unfortunately, these labels are not applied uniformly and leave large amounts of false content unmoderated.
This paper presents \systemname, a system that automatically identifies tweets that are candidates for soft moderation using Learning To Rank (LTR). 
We run \systemname on Twitter data to moderate false claims related to the 2020 US Election and find that it flags over 20 times more tweets than Twitter, with only 3.93\% false positives and 18.81\% false negatives, outperforming alternative state-of-the-art methods based on keyword extraction and semantic search.
Overall, \systemname assists human moderators in identifying and flagging false information on social media.
\end{abstract}

\section{Introduction}

The security research community has consistently been at the forefront of the fight against online abuse, from spam~\cite{grier2010spam,levchenko2011click}, phishing~\cite{ho2017detecting,zhang2021crawlphish}, to online fraud~\cite{christin2010dissecting,park2016understanding,suarez2019automatically}. 
Today, one of the most pressing types of abuse is the spread of false information, especially on social networks.
Arguably, mitigating it faces some unique challenges.
First, while some malicious actors spread misleading/false claims to advance their goals (\emph{``disinformation''}), false narratives are often believed by real users in good faith, who then re-share them on social media (\emph{``misinformation''})~\cite{krafft2020disinformation,saeed2021trollmagnifier,shao2018anatomy,singh2020first,starbird2019disinformation,vosoughi2018spread,zannettou2019let}. 
Second, identifying what is true or false is challenging, hard to automate, and often depends on external fact-checkers.
Finally, online platforms are often concerned about the effects of taking action on dis- and misinformation; 
for example, limiting what is allowed to be said on a platform can raise concerns about censorship and reduce engagement (and thus profit)~\cite{iglesias2021don,klonick2017new,myers2018censored}.
Nevertheless, the computer security research community is well poised to develop effective mitigation strategies for the problem of false online information, as highlighted by recent research in top tier venues in the field~\cite{mirzatactics,recabarren2022strategies,saeed2021trollmagnifier}.

As part of their mitigation strategy, social networks have begun to adopt so-called \emph{soft moderation}. 
Rather than removing content or banning accounts, they notify other users about false narratives and provide additional context.  %
Warning labels are attached to posts containing potentially false, misleading, or harmful claims, e.g., in the context of political disinformation~\cite{zannettou2021won} or COVID-19~\cite{ling2022learn,sharevski2021misinformation}.
An estimated 300,000 tweets were labeled under Twitter's Civic Integrity Policy~\cite{twitter_midterms_civic_info} as misleading around the 2020 US Elections, accounting for 0.2\% of all election-related tweets posted during that period.
Twitter reported that warning labels applied to tweets in late 2021 resulted in noticeable decreases in replies, retweets, and likes (13\%, 10\%, and 15\% reduction, respectively)~\cite{twitter_midterms_civic_info}.

\begin{figure}[t]
     \centering
     \begin{subfigure}[b]{0.4\textwidth}
         \centering
         \includegraphics[width=\textwidth]{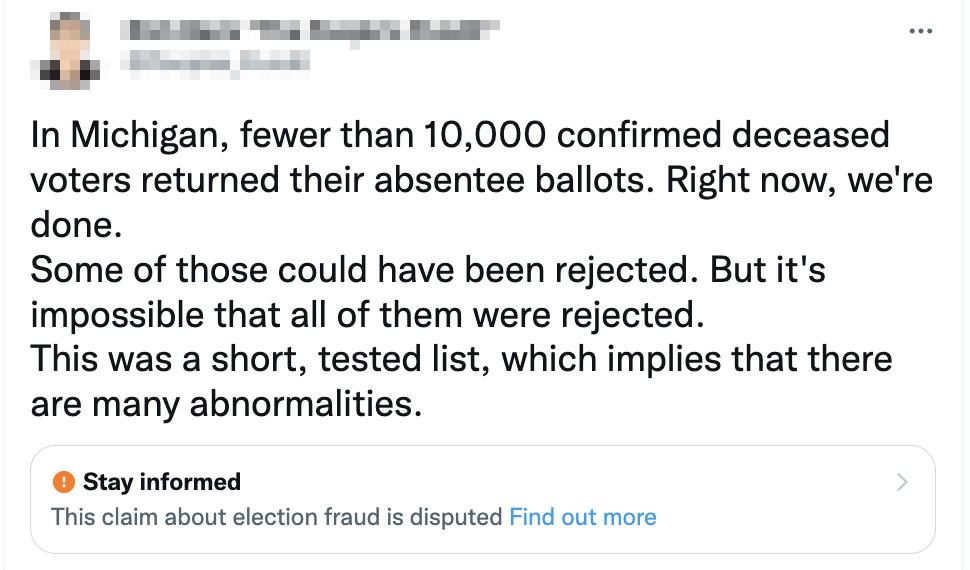}
         \caption{Moderated}
         \label{fig:tweet_moderated}
     \end{subfigure}
     \begin{subfigure}[b]{0.4\textwidth}
         \centering
         \includegraphics[width=\textwidth]{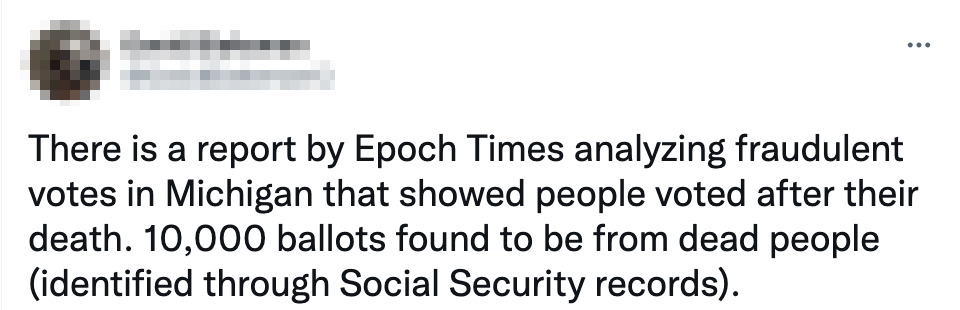}
         \caption{Unmoderated}
         \label{fig:tweet_unmoderated1}
     \end{subfigure}
\vspace{0.15cm}
        \caption{Two example tweets discussing alleged voting fraud in the State of Michigan during the 2020 US Presidential Election. Twitter added a warning label to the first but not the second one.}
        \label{fig:tweet_moderation}
\end{figure}

\descr{Motivation.} 
Early soft moderation results have been encouraging.
Prior work shows that warning labels may prompt site users to debunk false claims~\cite{zannettou2021won} or that they may reduce user interactions and extremism in comments~\cite{papakyriakopoulos2022impact}.
Unfortunately, details of Twitter's methodology are not publicly known.
Worse yet, recent work indicates that soft moderation may not be applied uniformly by Twitter, with ``benign'' tweets being erroneously labeled while misleading content goes unlabeled~\cite{lawfare_2022,zannettou2021won}.
For example, in Figure~\ref{fig:tweet_moderation}, we show two tweets discussing the same debunked narrative that fraudulent votes were cast for 10,000 deceased individuals in Michigan during the 2020 US Presidential election; one received the warning label, and one did not.

This example highlights the need for effective automated approaches to flag potentially misleading posts on social media.
Such approaches should cover as many misleading posts as possible while minimizing the number of unrelated posts that receive soft moderation labels to avoid \emph{alert fatigue} effects, where users start ignoring warnings if they become too frequent~\cite{ling2022learn}.
Recent work found that using overly-broad rules when applying soft-moderation labels (e.g., the inclusion of a specific hashtag) flags a large amount of unrelated content: 37\% of TikTok videos received COVID-19 related soft moderation were false positives~\cite{ling2022learn}.

\descr{Research Objectives.} In this paper, we set out to develop an automated system to flag candidate Twitter posts for soft moderation.
To minimize the false positive problem observed in previous work, instead of adopting a \emph{topic-specific} moderation approach (e.g., moderating any tweet containing a specific keyword), we follow a \emph{claim-specific} methodology, where moderation labels directly address the statement or claim made in the content they are applied to~\cite{morrow2022emerging}.
This fine-grained approach allows platforms to tailor warnings for specific claims, adjust severity cues on warning labels, and increase transparency by providing \emph{explainable} labels on why a specific tweet was moderated. %

\descr{Technical Roadmap.} We introduce \systemname, a system to assist in detecting candidate posts for soft moderation using a claim-specific approach.
\systemname takes as seed input a list of tweets moderated by Twitter and uses a Learning To Rank (LTR) based method~\cite{cao2007learning} to extract the optimal set of keywords that characterize the posts related to the same tweets. 
E.g., from tweets in Figure~\ref{fig:tweet_moderation}, \systemname would extract {\textit{``michigan,dead,ballot''}} as the best set of keywords to characterize discussion surrounding this claim in {\em both} tweets.
\systemname then uses these keywords to find more candidates for soft moderation.
We instantiate \systemname using the publicly available set of 2,224 tweets soft moderated by Twitter during the 2020 US Presidential Election, as curated by~\cite{zannettou2021won}.\footnote{https://github.com/zsavvas/Soft-Moderation-Interventions-Twitter}
\systemname retrieves the best set of keywords from 900 claims extracted from those tweets, identifying 2,042,173 additional candidates for soft moderation.

\descr{Main Contributions \& Findings.} 
We show that \systemname performs better than alternative approaches from keyword extraction and semantic search when recommending candidate tweets for moderation.
By manually analyzing the tweets flagged by our system, we find that our results are accurate, with 3.93\% false positives and 18.81\% false negatives, with both metrics being substantially lower than those reported by other approaches.
Moreover, \systemname reduces the number of tweets a human moderator would have to review by over five times compared to the second-best algorithm.
We also find that \systemname flags 20 times more candidates for moderation than those moderated by Twitter, suggesting that our approach can complement existing systems and improve the state of content moderation.

\systemname is platform-independent and can be bootstrapped from a single post corresponding to a narrative or an event deemed intervention-worthy by human moderators.
With the ability to scale to thousands of posts from a single misleading post, we show that event-driven keyword detection systems can be used as a foundation for large-scale soft moderation intervention systems.
At the same time, our results show that moderating content is a nuanced problem.
For example, posts discussing false narratives often attempt to debunk or ridicule them as satire.
Thus, we see \systemname as a tool to \emph{help} moderators identify potential candidates for soft moderation while leaving the final decision to humans.
We make \systemname's source code and the labeled dataset used in this paper publicly available.\footnote{https://github.com/idramalab/lambretta}

\section{Datasets}
\label{sec:datasets}

Our experiments use three datasets.
One includes the tweets that were soft-moderated by Twitter and is used as ground truth; the other two allow \systemname to find more tweets that are candidates for moderation in the wild.

\descr{Ground truth dataset.}
Our first dataset, denoted as \dtwo~, consists of the tweets that received soft moderation by Twitter released by Zannettou~\cite{zannettou2021won}.
It contains 2,244 tweets posted by 853 users.\footnote{Note that the dataset also contains 16,571 quoted tweets with warning labels; we exclude them since they are mostly used as additional commentary on the original tweet itself.}

\descr{Evaluation datasets.}
Two more datasets are used by \systemname %
to compile a set of tweets that are candidates for moderation.
One, denoted as \dthree~, is released by Abilov et al.~\cite{abilov2021voterfraud2020}; the other, denoted as \dfour~, is obtained by querying Twitter's Academic Research full-archive search endpoint.
\dthree\ contains 7.6M tweets related to the 2020 US Election.
More precisely, the authors of~\cite{abilov2021voterfraud2020} retrieve tweets from the Streaming API using a set of hashtags related to the voter fraud narrative surrounding the 2020 US Election (e.g., \#ballotfraud, \#voterfraud, \#electionfraud, \#stopthesteal).
The dataset is available as a list of tweet IDs.
After retrieving the complete tweet information from the Twitter API using these IDs we obtain 4,017,259 tweets.
Unlike \dthree~, \dfour~ is built at runtime leveraging Twitter's Academic Research full-archive search endpoint.
At various stages, we query this endpoint using keyword-based search, for example, as part of deriving features for the underlying ranking model.
The two different datasets %
capture two different types of data availability scenarios, and allow us to test \systemname on the entire Twitter archive using \dfour~.
Note that to match the time frame of our ground truth \dtwo~, we only extract tweets for the period between November 1 and December 31, 2020. 
In the rest of the paper, we refer to the evaluation datasets \dthree~ and \dfour~ as ``data store.'' 

\descr{Ethical Considerations.} Analyzing large-scale social media data may raise ethical concerns.
In this paper, we only collect public Twitter data using the official API and do not interact with users.
As such, this work is not considered human subjects research by our institution.
Regardless, we are mindful of the privacy of Twitter users and do not analyze any personally identifiable information (e.g., location data, account names, etc.).
Also, when presenting example tweets in this paper, we apply ``heavy disguise'' and paraphrase them to make user re-identification more difficult~\cite{fiesler2018participant}.

\section{Overview of \systemname}

\begin{figure*}
 \includegraphics[width=0.999\textwidth]{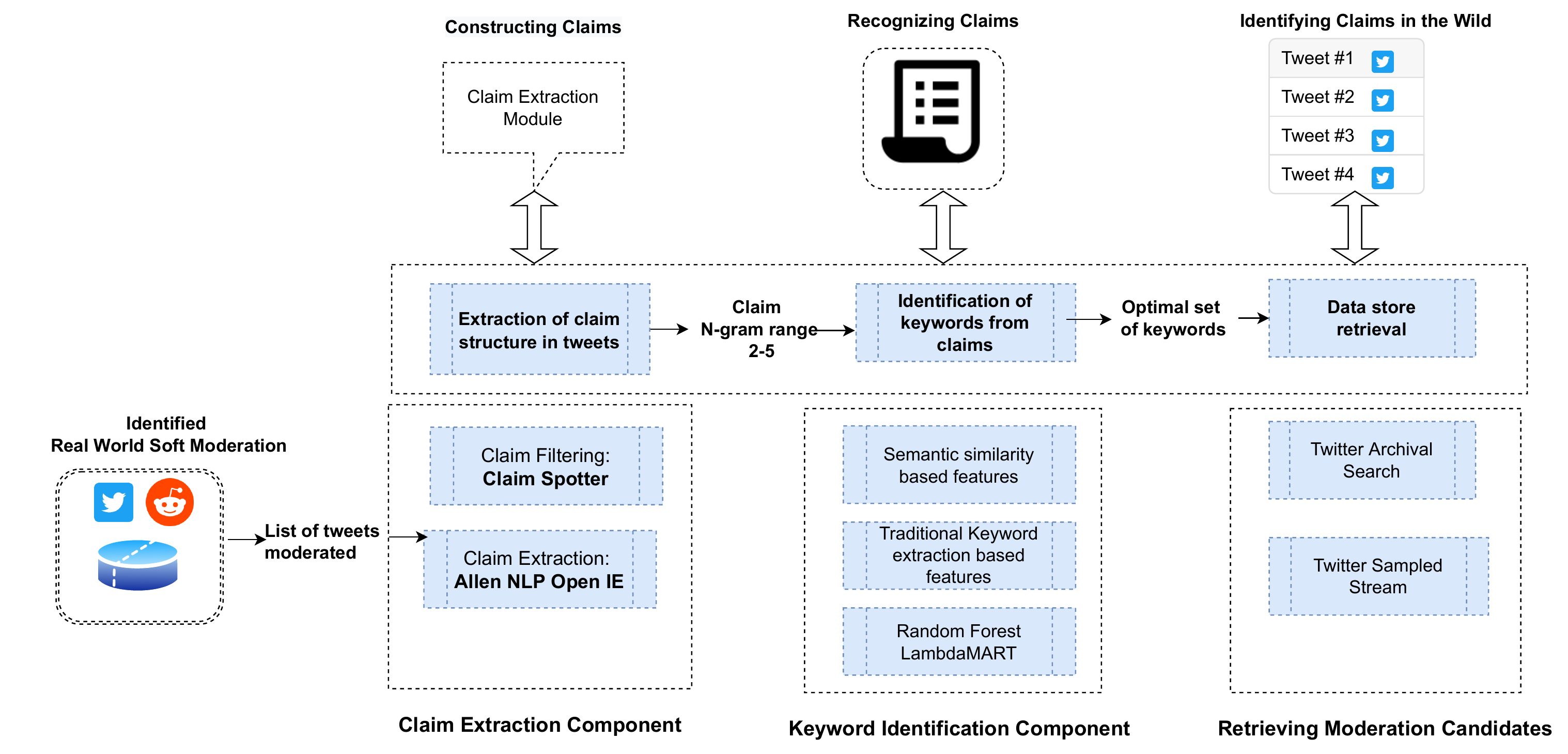}
 \caption{System overview of \systemname.}
 \label{fig:sysdiagram}
 \end{figure*}

This section presents the different components of our system and how the end-to-end pipeline operates.
\systemname takes a seed list of tweets
discussing a false claim and identifies similar candidate tweets for moderation.
A high-level overview of our system is presented in Figure~\ref{fig:sysdiagram}.
\systemname has three stages: 1)~extracting claim structures from tweets,
2)~training a Learning to Rank (LTR) model from claim structures to extract the most relevant set of keywords for a given claim,
3)~identifying candidate tweets from the data store similar to the seed tweets for platform moderators to make further decisions on.

\subsection{Claim structure extraction}
\label{sec:claimextraction}

\systemname extracts claims that should be moderated, starting from one or more tweets.
These extracted claims have similar structures to those manually curated by fact-checking organizations like Snopes\footnote{https://www.snopes.com/fact-check/} or PolitiFact.\footnote{https://www.politifact.com/factchecks/}
Automatically extracting claims from tweets is useful for our purposes for two reasons.
First, it narrows down spans of tweets containing assertions and provides a better training dataset for \systemname's subsequent phases.
Second, while we could directly rely on the claims published by fact checkers instead of building them ourselves, this is not ideal since fact checks tend to lag behind the appearance of false narratives by 10-20 hours~\cite{shao2016hoaxy}, which impairs real-time moderation.

To identify claims in the seed list of tweets, we follow an approach inspired by linguistics.
First, we identify propositions in tweets, defined as ``a declarative piece of text used to make a statement or assertion~\citep{palau2009argumentation}.''
However, not all linguistic propositions contain a claim that should be part of soft moderation efforts.
To account for this, \systemname runs identified propositions through a claim classification component. 
The output of this phase is a set of claims that can be passed to \systemname's later stages.

\descr{Extracting propositions.}
As an operational definition, we consider a proposition as a standalone span of the tweet, potentially containing a claim~\citep{palau2009argumentation}.
Not all propositions are a claim, despite being a statement or assertion.
We will discuss the detailed specifications of what constitutes a claim later. 

It is important to note that one tweet might contain multiple sentences, and a single sentence might contain multiple propositions.
We draw motivations from prior work on claim extraction on Twitter~\citep{lim2016claimfinder}, where the authors formulate the problem by using propositions extracted through Open Information Extraction (Open IE).
The major advantage of using Open IE is the ability to extract an unbounded number of propositions in a sentence without requiring any domain knowledge of the underlying text, making it usable off-the-shelf.
In a similar problem setting~\citep{lim2016claimfinder}, an Open IE system called Clausie~\citep{del2013clausie} efficiently clustered tweets related to two real-world events into twelve different claims.
In our implementation of \systemname, we use a tool developed by Stanovsky et al.~\citep{stanovsky2018supervised}, which frames Open IE as a sequence tagging problem and uses a bi-LSTM transducer, available via Allen NLP~\citep{gardner2018allennlp}.
For the rest of the paper, we refer to this Allen NLP Open IE extractor as the ``Proposition Extractor.''
As an example, using the Proposition Extractor on a tweet with the body ``Mail-in voting eases access barriers that might otherwise exclude voters physically unable to cast votes and has shown increased turnout across demographic groups.'' extracts the following sets of propositions: 1)~``Mail-in voting intended to ease access barriers'', 2)~``access barriers that might otherwise exclude voters physically unable to cast votes,'' and 3)~``Mail-in voting has shown increased turnout across demographic groups.''

\descr{Building ground truth annotations.}
As a natural first step, we need to validate the applicability of the Proposition Extractor in extracting potential \emph{claim} structures.
We thus sample 200 random tweets from \done~ and annotate spans that contain a claim.
Before annotating spans in a tweet containing claims, we need to formally define the scope of a claim. %
The lead author manually annotates our 200 tweet sample by adapting the scheme proposed in~\citep{konstantinovskiy2021toward}, which identified six different categories of claims via crowdsourced annotations of sentences extracted from subtitles of four UK political TV shows.
We consider a tweet as containing a claim if the tweet:
\begin{compactitem}
\item discusses an event occurring in a particular US state related to the 2020 elections;
\item makes an assertion that can be further validated or debunked;
\item cites events and stories related to election manipulation;
\item references statistics to support or potentially deny an argument for election manipulation;
\item reports breaking news or a developing event related to the 2020 election; or
\item mentions actors often linked to conspiracy theories (Soros, Clinton, etc.) in relation to election events.
\end{compactitem}
In the end, we find 115 tweets with at least a single claim in our random sample of 200.

To annotate claim \emph{spans}, which are normalized units of claims embedded in a tweet, we follow the method outlined by Wuhrl and Klinger, who perform claim detection on biomedical tweets related to topics like COVID-19, measles, and depression~\cite{wuhrl2021claim}.
They present a guideline to human annotators on the task of annotating claim spans within a tweet.
From the guidelines, claim spans are the central statement of an argumentative structure, phrased with an argumentative intent while expressing a clear stance (support or oppose) about the argument. 

This results in a ground truth dataset of 144 claim spans for the 115 tweets that contain a claim.
To exemplify the type of claims we identified, we present the following three tweets, with the claim spans in bold.
\smallskip
\begin{mdframed}[style=MyFrame,nobreak=true]
\begin{quote}
\small
\begin{itemize}[leftmargin=*]
\item PA election night narrow lead was erased by \textbf{hundreds of thousands of mail-in ballots counted without Republican observer.} I wonder why were Republican observers excluded from the counting?
\item Sidney Powell ( part of Giuliani's team ) is explaining that the \textbf{ plan to steal the election from Trump was Hugo Chavez}. However, Chavez died in March of 2013. 
\item A September 2020 study released by Judicial Watch revealed that \textbf{U.S. counties had 1.8M more registered voters than eligible} voting-age citizens. That means that in these counties the \textbf{registration rate exceeded 100\% of eligible voters.}
\end{itemize}
\end{quote}
\end{mdframed}

\noindent Some examples of tweets that do not contain a claim are:
\smallskip
\begin{mdframed}[style=MyFrame,nobreak=true]
\begin{quote}
\small
\begin{itemize}[leftmargin=*]

\item ``Republicans need to rise together and stand against the fraud. Otherwise, our party might never win election again.'

\item ``Dems thought they would get away with cheating. But they got caught. Funny enough how quiet they are? I would fight really hard to clear my name if I was accused of cheating. Silence speaks louder than words.''

\end{itemize}
\end{quote}
\end{mdframed}

We test the Proposition Extractor's ability to recognize all potential claim structures present in a tweet as propositions.
First, we split the 200 tweets using a sentence tokenizer and query each sentence against the Proposition Extractor.
We then compare the set of propositions returned by Proposition Extractor against the annotated ground truth spans, looking for missed ground truth spans.
These types of errors are called \emph{uninformative extractions}~\citep{etzioni2011open}, where Open IE systems can omit critical information from the sentences in the propositions they identify.
The Proposition Extractor extracts 264 propositions from 115 tweets containing claims, and 159 propositions from 85 tweets without claims.
Of the 144 claim spans in our ground truth, the Proposition Extractor
 misses only four instances.

\descr{Filtering propositions that are claims.}
Thus far, we have 264 propositions extracted from 115 tweets, and only 144 of these propositions are annotated as claims.
As discussed above, a tweet containing a claim span can have multiple other propositions that are not a claim, and we need to filter out these too.
As a final step, \systemname needs to distinguish propositions that are claims from those that are not. %
We address this by further scoring the propositions extracted by the Proposition Extractor against a state-of-the-art claim classifier called ClaimSpotter~\citep{claimspotter}, which leverages a gradient-based adversarial training on transformer networks to identify claims and is trained on manually labeled sentences from historical US presidential debates~\citep{meng2020gradient}.
For any input text, ClaimSpotter returns a score between 0 and 1, representing how ``check-worthy'' a claim is.
This is not how likely something is to be \emph{true}, but instead if it makes an objective claim (1) vs. espousing a subjective opinion (0).
For instance, a relatively subjective proposition such as \textit{``And that, ladies and gentlemen, is how you steal an election''} returns a score of \textbf{0.285}, whereas a relatively objective proposition such as \textit{``In the past 20 years there have been approx 250 million votes cast and around 1200 proven cases of voter fraud"} scores \textbf{0.85}. %

While the ClaimSpotter API is already trained and publicly available, we need to devise an appropriate threshold that gives us accurate results for classification considering our problem setting. 
To this end, we obtain the Claim Spotter scores for all 144 claim spans annotated as ground truth and set the target class for these scores as 1.
Similarly, we obtain the ClaimSpotter scores for the 159 propositions extracted from the 85 tweets without claims and the 120 propositions extracted from the 115 tweets with claims.
We first perform a 75--25 train-test split to identify the optimal threshold of 0.490 based on the Receiver Operating Characteristic (ROC) curve.
We then perform a 5-fold cross validatio- using this threshold, obtaining an average Precision of 0.881 and an average Recall of 0.895.
The system misclassifies claim propositions as non-claim 3.46\% of the time, while non-claim propositions are classified as claims with a rate of 6.58\%.
We deem these results acceptable and thus adopt 0.490 as our threshold for \systemname's claim extraction component.

\descr{Extracting claims from the entire dataset.}
From the remaining 2,044 tweets in \done~ we extract 4,471 propositions, 756 of which are identified as claims based on our decision threshold. 
In addition to the 144 claims identified during the decision threshold calibration phase, this brings us to 900 total claims extracted from 2,244 tweets in \done~.
Some examples of tweets that contain claims include:

\smallskip
\begin{mdframed}[style=MyFrame,nobreak=true]
\begin{quote}
\small
\begin{itemize}[leftmargin=*]
\item ``The voting machine in Green Bay ran out of ink, which delayed the final results. An election official went back to City Hall to get more ink.''
\item ``The Dominion Voting Systems machines switched over 6,000 ballots for Biden in Michigan. 
They are used throughout the United States on a wide scale in sixteen states. Virtually all of these states are blue states. 
The system needs to be audited to make sure the right person is elected.''
\end{itemize} 
\end{quote}

\end{mdframed}

These extracted claims are topically diverse, covering different aspects of the election process as multiple phases of the election event developed.
We find many claims discussing mail-in ballots and ballot harvesting posed as a threat (e.g., \textit{``poll workers stuffing ballot box with mail-in ballots for Democrats''}) %
and discussion of election fraud caused by voting machines (e.g., \textit{``Dominion machine flipped votes from Trump to Biden''}). %
Additionally, we find many claims misinterpreting bipartisan events of ballot counting as partisan vote counting (e.g., \textit{``spike for Biden votes as suitcases of ballots started to be scanned''}) %
and presence of dead, fake, and ineligible voters (e.g., \textit{``86,845 mail-in ballots lost in Arizona''}). 
Finally, many claims in our dataset discuss the developing situation of the vote totals as statistically suspicious or rigged (e.g., \textit{``mysterious spike of votes in Pennsylvania had 600,000 votes for Biden and only 3,200 for Trump''}). %
This diversity of topics found in the claims will be an important evaluation setup for \systemname and its ability to generalize over various topics about an event.  

\subsection{Identification of keywords from claims}
\label{sec:keyword}

After summarizing tweets into claims, \systemname extracts the most representative set of keywords from claims.
We need to further extract keywords from our claims because querying Twitter search with the full claim text will give us a very limited set of tweets discussing the claim in a narrow way.
The following are three tweets discussing the same claim related to voting machines:

\begin{mdframed}[style=MyFrame,nobreak=true]
\begin{quote} 
\small
\textbf{Claim:} ``smartmatic foreign software voting machine design rig election socialist venezuela.''

\textbf{Example Tweets discussing the same claim:}
\begin{itemize}[leftmargin=*]

\item ``Dominion developed by Sequoia-Smartmatic, a company previously owned by Chavez. 
US Intel said Smartmatic was used to rig the 2004 election in Venezuela.
The Chicago election commission concluded the Smartmatic software delivers the results desired to the election officials.''

\item ``China, Cuba  Venezuela, all these communist countries and Antifa interfered in U.S elections. They used Smartmatic voting software, created by Hugo Chavez. Guess who helped them :  George Soros and the Clinton Foundation, rigging U.S election for Joe Biden.''

\item ``The Dominion Smartmatic machine was used in Venezuela by Cuban intelligence in an attack on its democracy by Chavez and Castro.
Warfare through Voting systems has replaced guerrilla  attacks against nations.''
\end{itemize}
\end{quote}
\end{mdframed}
The first tweet has a very formal tone to it, is detailed, and uses some sources (US Intel, Chicago Election Commission) to establish context.
On the other hand, the second tweet also talks about Smartmatic and its connection with Venezuela, but has totally different actors surrounding the discussion (Antifa, George Soros, and the Clinton Foundation).
Finally, the third tweet is more of a socio-political commentary espousing the misleading claim.
These examples illustrate that it is important to extract the most representative set of keywords for any claim to understand the wide variety of ways it can be discussed.

\descr{Problem with existing Keyword Identification techniques.}
Several approaches to extract important keywords from social network posts have been proposed~\cite{grootendorst2020keybert,campos2018yake,mihalcea2004textrank,lee2009keyword}.
We initially experimented with various existing keyword detection methods~\cite{grootendorst2020keybert,campos2018yake} to extract the best set of keywords from the claims extracted in \systemname's previous step.
Our analysis identified three main issues with previous approaches.
We include three examples below obtained while testing with Yake~\cite{campos2018yake}:
\begin{mdframed}[style=MyFrame,nobreak=true]
\begin{quote}
\small
\begin{itemize}[leftmargin=*]
\item Problem Type 1: Missing key entities

\textbf{Claim: Michigan Governor Whitmer send health dept into Detroit TFC Center to evict GOP poll-watchers but not Dem pollwatchers}

Automatically detected keywords using YAKE:  \textit{Detroit, TFC ,pollwatcher}

\item Problem Type 2: High Recall, Low Precision

\textbf{Claim: Signature verification system in Clark County have 89\% failure rates for catching poor signature matches}

Automatically detected keywords using YAKE:  \textit{signature, match}

\item Problem Type 3: Low Recall, High Precision

\textbf{Claim: Tens of thousands of votes illegally received after 8 P.M. on Tuesday, Election Day}

Automatically detected keywords using YAKE:  \textit{ten,thousand,vote,8}

\end{itemize}
\end{quote}
\end{mdframed}

In the first example, the keyword extractor misses the key entities of the claim being discussed, \textit{Whitmer} and \textit{GOP}, and may produce many unrelated results.
In the second example, the extracted keywords do capture the key entities, \textit{signature} and \textit{match}, but these keywords are generic and may result in many hits that are part of the larger narrative around signature matching on mail-in ballots rather than the specific claim of the failure rate of signature matches in Clark County.
Finally, the third example extracts overly specific keywords that will miss relevant content.
The issue with existing approaches is that they identify the most important keywords by only looking at the claim body (e.g., the tweet already moderated by Twitter), without taking into account the context in which the claim is discussed.
To overcome this issue, \systemname formulates the problem as a Learning to Rank (LTR) task~\cite{cao2007learning}, where the selection of optimal keywords from the claim is driven by the document store (i.e.,  \dthree~, \dfour~, etc.) containing examples of social media discussion of the claim in question.
We note that we use both \dthree~ and \dfour~ in our experimental setup to cross-evaluate the LTR model trained on one document store and tested on an unseen document store. 
We provide more details on the LTR task and our keyword detection experiment below.

\descr{Learning to Rank.}
Learning to Rank (LTR) is a task aiming to learn a function to rank the effectiveness of query terms in retrieving relevant documents from a document store.
LTR based methods have been used in many information retrieval oriented applications~\cite{liu2009learning,surdeanu2008learning,shaw2013learning,leaman2013dnorm,karatzoglou2013learning}.
We refer readers to~\cite{cao2007learning} for a detailed review of the LTR task and different approaches to it.

To train our LTR model, we first annotate ground truth keywords for a fraction of the 900 misleading claims extracted from \dtwo~, manually labeling relevant tweets resulting from querying the keywords.
We then develop features to train the LTR model and evaluate it on unseen tweets from \dthree~ and \dfour~.
We describe these steps in detail below.

\descr{Building ground truth annotations.}
We begin by randomly selecting 125 misleading claims from our filtered set of claims (see Section~\ref{sec:claimextraction}) to train our LTR model.
We refer to these claims as \initialtrainclaims~.
For each claim, we annotate the ground truth to be the set of keywords made up of terms from the misleading claims which produce the most related set of results when queried against our data store, optimizing for both relevance and size of results.

We start by initializing \basekeyword~, which are two words consisting of the subject and the object of a misleading claim.
We query the two different Twitter data stores \dthree~ and \dfour~  with the \basekeyword~ and retrieve a set of results.
As expected, the \basekeyword~ are usually pretty broad and return many irrelevant false positives.
We fine tune the query by checking a random sample of 20 posts from query results and then adjust the \basekeyword~ by either adding new words from the query or removing words that are causing the false positives.
We repeat the process until we find the most relevant set of keywords for each misleading claim.
Finally, for each claim, the best set of keywords is tagged as a positive instance with the remaining keywords tagged as negative instances of relevance.
This ground truth is then used by \systemname to learn the ranking function that automatically identifies the best set of keywords for any given claim.
This enables \systemname to automatically extract keywords from the body of a claim without any further external context of human intervention.

\descr{Data pre-processing.}
Before further analysis, \systemname removes stopwords from the tweet claims while doing basic pre-processing, e.g., lowercasing text and removing punctuation marks.
We keep numbers since they can often be an integral part of extracted claims.
We split the pre-processed misleading claims into n-grams of length two, three, four, and five, which are our potential set of query terms.

\descr{Feature Engineering.}
Next, we use the potential query terms returned by the previous step to extract our learning to rank model features.
To this end, we query the two data stores \dthree~ and \dfour~, retrieving all the posts matching the query terms. %
LTR requires us to generate a dataset consisting of a query set, relevance information, and feature values to learn the ranking.
Feature values can be a few or numerous.
Applications of LTR have used features like document TF-IDF, BM25~\citep{ramos2003using} scores, document length, number of matching query terms, and number of query terms in important sections of a document, e.g., the title of a Web page~\cite{macdonald2012usefulness,bai2010learning,surdeanu2011learning}.
The most widely used benchmark to build models based on LTR is the LETOR dataset~\cite{qin2010letor}, which contains query sets, learning features, and labeled rankings related to the 25 million page \textit{GOV2} Web page collection~\citep{qin2013introducing}. %
Unfortunately, we cannot directly use the LETOR dataset since our data store is composed of posts from Twitter, which are fundamentally different from Web pages.
Instead, we take inspiration from the LETOR dataset and develop six features that we use in our LTR experiments.
In the following, we briefly discuss these features.

We use the term \spanningsubset~ to describe the earliest 20\%, most recent 20\%, and 10\% of the middle-aged results returned from the query, based on the timestamp attached to their tweets.
From this, we derive six features:
\begin{enumerate}
\item Total number of hits (matching tweets) produced.
\item Mean and median pairwise similarity score between the entries of the \spanningsubset~.
\item Mean and median similarity score between the entries in \spanningsubset~ and the claim.
\item Mean and median similarity score between the query and the claim.
\item Mean and median value of the TextRank scores~\cite{mihalcea2004textrank} of the query terms.
\item Mean and median score of the Term Frequency-Inverse Document Frequency scores~\cite{ramos2003using} of the query terms.
\end{enumerate}

We need to extract \spanningsubset~ from the set of returned results as the number of results retrieved by the candidate keywords from the candidate query set might grow large, specially in cases of generic set of keywords in a query.
Performing a pairwise semantic similarity comparison on this large set is not computationally feasible.
The intuition behind \spanningsubset~ is to sample tweets during the different period of a discussion (from the onset to the current phase) along a timeline, compared to randomly sampling tweets.
The similarity score between the set of results, and between the query and the results are calculated using the cosine similarity of the sentence embeddings encoded using a pre-trained \textit{all-mpnet-base-v2} model proposed in~\citep{song2020mpnet}.
The \textit{all-mpnet-base-v2} model produces sentence embeddings with 768 dimensions, and had the best average performance on encoding sentences over 14 diverse tasks from different domains.
Note that the features used in training the LTR model are not domain-dependent (i.e., election misinformation in this case), and are designed to entirely capture the semantic relatedness between query results and subsequent claims.
These type of features will be useful for applying \systemname in other contexts for content moderation in different topics.

\descr{Training the LTR model.}
We use the RankLib project, part of the Lemur Toolkit~\cite{lemurtoolkit} which includes a variety of LTR algorithms.
While there are LTR algorithms that use complex neural architectures like Deep Learning~\cite{pasumarthi2019tf,chen2020deeprank}, we cannot directly use them since they are designed to work on large scale datasets like LETOR.
Instead, our LTR models are powered by features utilizing state of the art neural semantic models (\textit{all-mpnet-base-v2}) to capture the interaction between query terms and result set.

We use all eight algorithms implemented by Lemur (MART, RankNet, RankBoost, AdaRank, Coordinate Ascent, LambdaMART, ListNet, and Random Forests) for our experiments.
To evaluate our model, we use a rank-based evaluation metric commonly used in information retrieval settings: Mean Average Precision (MAP).
In our case, our problem setting is a binary judgement where a keyword is either relevant to the misleading claim or not.

As with all keyword extraction problems, our dataset has many irrelevant combinations (majority class) of wrong keywords compared to one or two correct combinations (minority class) of optimal keywords we want to extract.
This imbalance causes the learning algorithm to perform very poorly.
To address this problem, prior work on LTR used BM25 as a pre-ranker to retrieve a small set of highly ranked documents from the entire document index before applying the LTR algorithm on the small subset of highly relevant documents~\citep{liu2009learning}.
We employ a similar pre-ranking step to filter out the set of keywords that retrieve irrelevant results for the claim by using a heuristic based on semantic query similarity.
The idea behind the heuristic is that candidate keywords most likely to be optimal return tweets that are more semantically similar than unrelated ones.
Thus, for each claim, its corresponding set of candidate keywords, and the retrieved results using these keywords as queries, we construct a subset called \textbf{FilteredQuerySet}, which is a ranked list of the top \textit{k} results retrieved by all candidate keywords, sorted by the cosine similarity with the claim under question.
The results returned by an irrelevant query have lower semantic similarity with the claims, thus failing to appear in the ranked set \textbf{FilteredQuerySet}.
We experiment on different values of \textit{k} and find that setting \textit{k} to 20 includes all of the ground truth queries for training, validation, and test claims, while reducing the size of irrelevant candidate query set by 30 times.

To test our LTR model, we perform a 5-fold cross-validation on our ground truth.
We observed that Random Forest with LambdaMART~\cite{burges2010ranknet} as the bagging ranker produced the best results among the eight different ranking algorithms.
We further increase performance by performing a grid search over hyper-parameters (number of leaves, number of bags, number of trees, and minimum leaf support).
The 5-fold cross-validation on our ground truth using the tuned Random Forest model achieves a MAP of 0.768.
As we will show later in Section~\ref{section:comparison_model}, this convincingly
outperforms other state-of-the-art keyword extraction approaches.
We further refer to this LTR model as \initialtrainmodel~.

\descr{Validating the LTR model.}
The previous experiment showed that our LTR approach can produce an accurate model on our training set.
We now want to understand whether our model generalizes and can effectively identify posts related to claims that are not in the training set.
To this end, we design and conduct two experiments.
In the first experiment, we randomly select 75 additional misleading claims (extracted as per Section~\ref{sec:claimextraction}) not part of the ground truth set, referring to them as \validationclaims~.
The output keywords for these new claims can be inferred from the previously trained model, but they still require ground truth annotation to evaluate results.
We thus manually label the \validationclaims~ via the same iterative method that we used for the \initialtrainclaims~.

Next, we generate the potential query terms set for the \validationclaims~, following the same steps for \initialtrainclaims~.
We then query \dthree~ and generate the feature values for candidate keywords for each claim.
Finally, we perform inference on these new claims to see if our trained model can identify the best set of keywords.
Our model achieves a MAP of 0.781, indicating that our approach effectively identifies keywords and retrieves data for previously unseen claims.
After the validation step, we now have ground truth of 75 additional misleading claims from \validationclaims~, which we use to expand our overall training set to a total of 200 claims, which we call \expandedclaims~.
We train a new LTR model on the \expandedclaims~, which we refer to as \expandedmodel~.
At this point, we have 700 claims remaining which were not manually annotated and are missing corresponding keywords.
We later use the \expandedmodel~ to extract the keywords for these claims.

In the second experiment, we aim to verify that the learned model is not biased towards the data store it was trained on (i.e., \dthree~) and can be applied to an unseen data store (\dfour~).
In the previous experiments we generated the features from results queried on \dthree~ for validation.
Instead, in this experiment we build the features from results queried on \dfour~.
We re-use the \expandedclaims~ as our claim list, along with the corresponding ground truth we had collected for the previous training/validation experiments.
Following the same steps as for the \initialtrainclaims~ and \validationclaims~, we generate the feature values for each of the candidate keywords for the claim by querying the datastore of \dfour~.
We use the previously trained \expandedmodel~ model for inference on the \expandedclaims~ and achieve a MAP of 0.767, showing that our LTR model is not dependent on the datastore it was trained on.

\subsection{Data store retrieval}
After training our LTR model and validating its performance on ranking experiments with other methods, we apply the ranking model to the remaining 700 misleading claims.
The output is 499 unique sets of keywords. 
This number is lower than the total number of claims (900) since the keywords extracted from two different claims can be the same.
We use these 499 sets of keywords to query the two data stores \dthree~ and \dfour~.
We require candidates to match only if all keywords in a query are present in a tweet, regardless of the order of the tokens.
We also make sure to exclude retweets and quoted tweets when searching for the relevant tweets.

After searching for the keywords, we obtain 2,042,173 tweets from \dfour~ and 101,353 tweets from \dthree~.
The average number of tweets per misleading claim from \dfour~ is 5,988 and for \dthree~ it is 203.11.
We use the 101,353 tweets from \dthree~ to further evaluate \systemname throughout the rest of the paper, as checking the tweets flagged for moderation from \dfour~ is not feasible due to Twitter API limitations.

\section{Evaluation}
\label{sec:evaluation}

To evaluate \systemname, we first compare the quality of the candidate tweets extracted by our system to those recommended by other state-of-the-art approaches. %
Next, we manually analyze a subset of the tweets flagged by \systemname to assess its false positives and false negatives.
We then check whether the tweets flagged by \systemname received soft moderation from Twitter, in the context of the 2020 US Election voter fraud allegations, finding that only a small fraction of them did.
Finally, given the disparity between the soft moderation candidates flagged by \systemname and those that received labels by Twitter, we use our dataset to understand if Twitter's moderation is driven by specific characteristics of the tweets or of the users posting them.

\subsection{Comparison with other keyword extraction and information retrieval based methods}
\label{section:comparison_model}

In Section~\ref{sec:keyword}, we showed some examples of why existing keyword extraction methods might be unsuitable for our task, motivating us to develop the LTR component of \systemname.
We now provide a rigorous quantitative analysis of this fact, by comparing the LTR model used by \systemname with three other keyword extraction algorithms: YAKE~\cite{campos2018yake}, KeyBERT~\cite{grootendorst2020keybert}, and RAKE~\cite{rose2010automatic}. 
As an alternative to keyword extraction algorithms, another possible approach to finding similar tweets given a source one is leveraging semantic search techniques~\citep{guha2003semantic}.
We also evaluate state-of-the-art methods in this space against \systemname's LTR model; more precisely semantic search using Sentence Transformers~\citep{reimers-2019-sentence-bert} and BM25~\citep{bm25}, which can be used to get tweets matching a query tweet by using a ranking function.

To establish ground truth, we sample 60 random claims from the set of 200 \expandedclaims~. 
For these sets of claims and the ground-truth keywords, we manually verify that each tweet returned by the keywords does discuss the claims in question and filter out any irrelevant ones.
This yields a set of 10,776 tweets associated with the 60 claims. %
Sentence Transformers and BM25 also require tuning a similarity threshold for matching and ranking tweets; for these experiments, we retrieve tweets using different thresholds, ranging from 0.3 to 0.9, and select the threshold for each method that achieves the highest F1 score based on our ground truth. 

\begin{figure}[t]
     \centering
         \includegraphics[width=0.99\columnwidth]{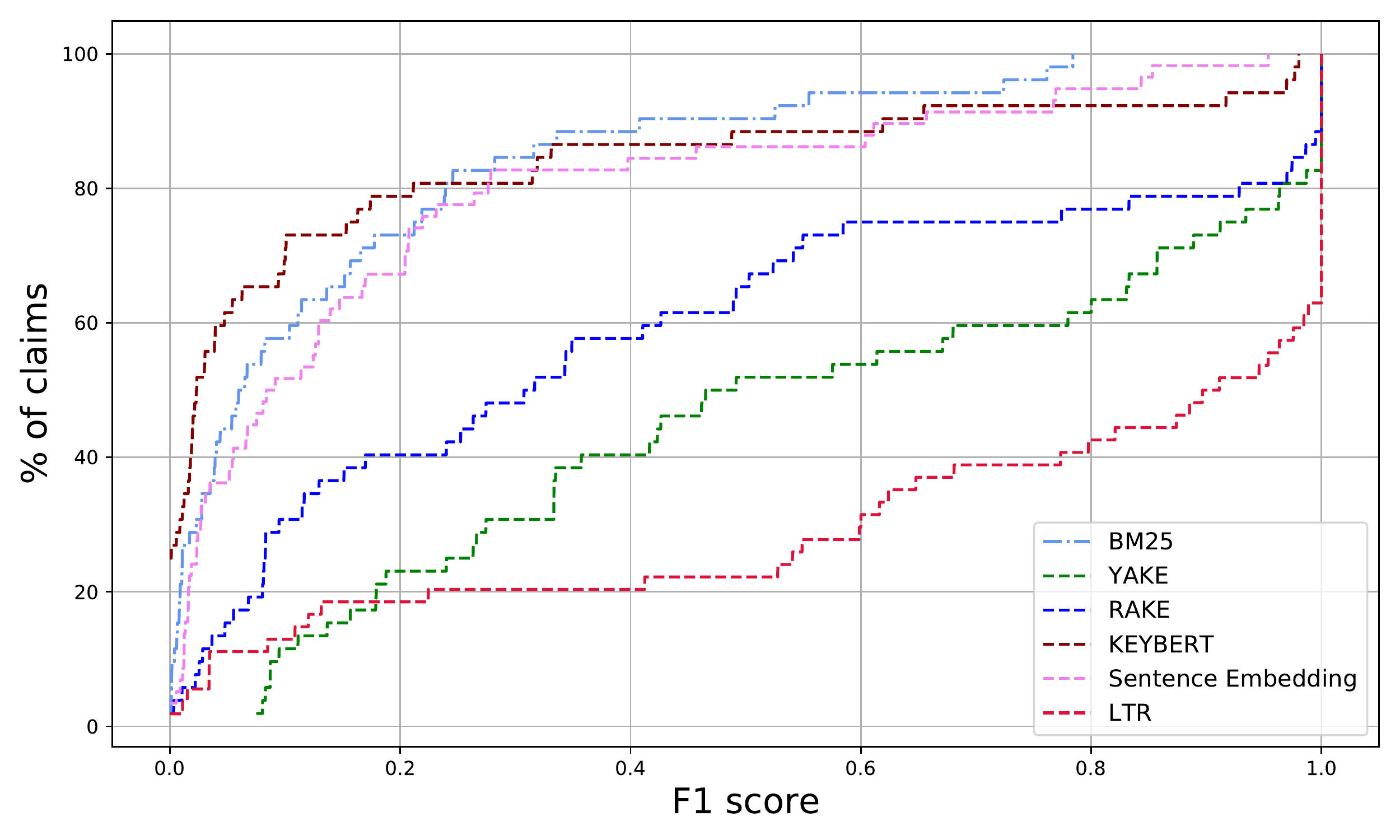}
	 \caption{Performance of the LTR model vs. other keyword identification and semantic search methods.}
         \label{fig:keywordvalidation}
\end{figure}

Figure~\ref{fig:keywordvalidation} reports the F1 score for all methods, including LTR, when extracting similar tweets to the 60 claims curated from \expandedclaims~.
\systemname's LTR is the best performing model, with over 60\% of the claims having an F1 score of 0.8 or higher.
The second best performing algorithm is YAKE, with less than 40\% of claims having an F1 score of at least 0.8.
BM25 and semantic search using Sentence embeddings perform poorly, as does KeyBERT.
These results confirm that the LTR model, the keyword identification component of \systemname, is effective at identifying similar tweets with better Precision and Recall than other keyword extraction and semantic search based methods. 

\descr{Comparison of the reduction of effort offered by different methods to content moderators.}
The goal of \systemname is to provide a set of social network messages to content moderators, allowing them to make informed decisions and to keep the platform safe.
To be useful and avoid overwhelming content moderators, this system should reduce the number of candidates as much as possible, while at the same time maximizing Precision and Recall.
From our comparison experiments described above, we find that \systemname reduces the pool of tweets which should be checked for moderation compared to all other systems.
\systemname retrieves 5.33 times less tweets related to a moderated claim than the second best performing system, YAKE.
This shows that \systemname is better suited than alternative approaches for the task of aiding moderation of misleading information on social media.

\subsection{Validation of \systemname}
\label{sec:validation}

In this section, we first perform a manual analysis on the results recommended by \systemname, assessing its False Positives and False Negatives.
We then check if the recommendations made by \systemname were also soft moderated by Twitter, finding that only a small fraction of tweets flagged by \systemname were intervened by the platform.

\descr{Manual validation.}
To understand the quality of the recommendations made by our approach, we perform a manual examination of the tweets recommended by \systemname, aiming to check if they should indeed have been moderated.
The first author of this paper samples 1,500 tweets among the candidates flagged by \systemname and analyzes them qualitatively, identifying seven categories that they can fall under: 1)~Amplyfing tweets, 2)~Reporting Tweets, 3)~Counter Tweets, 4) Satire Tweets, 5)~Discussion Tweets, 6)~Inquiry Tweets, 7)~Irrelevant Tweets.
The process goes through multiple iterations of coding the sample tweets, grouping them into different categories. 
The annotator then consolidates overlapping themes and categories, converging to seven which we discuss below.
Note that a single tweet can fall under multiple categories, e.g., they can amplify a misleading claim and prompt a discussion at the same time.
The definition of each category, alongside a representative example tweet of the category is presented below:
\descrit{1) Amplifying Tweet:} it positively reinforces the misleading claim and aims to further spreading the message.
\begin{mdframed}[style=MyFrame,nobreak=true]
\begin{quote}
\small

``Terry Mathis (born 1900) apparently voted via absentee ballot in Wayne County: Michigan.  It doesn't stop here. This person applied for an absentee ballot on December 2: the ballot was then sent out AND returned in the same day.<URL> <URL>''
\end{quote}
\end{mdframed}

\descrit{2) News Reporting:} it reports the headline of a misleading news article or another tweet, without any additional commentary and text from the tweet's author.
\begin{mdframed}[style=MyFrame,nobreak=true]
\begin{quote}
\small
``BREAKING: Unofficial: Trump trailing Biden by only 4,202! There is a ballot count upload glitch in Arizona. Reports saying over 6,000 False Biden Votes Discovered <URL>.''

\end{quote}
\end{mdframed}

\descrit{3) Counter Claim:} it attempts to question and/or debunk the misleading information.
\begin{mdframed}[style=MyFrame,nobreak=true]
\begin{quote}
\small
``Misleading claims that Trump ballots in Arizona were thrown out because Sharpie pens were provided to voters are untrue. A ballot that cannot be read by the machine would be re-examined by hand and not invalidated if it was marked with a Sharpie.\#Election2020 <URL>.''

\end{quote}
\end{mdframed}

\descrit{4) Satire:} it discusses the false claim in a satirical way.

\smallskip
\begin{mdframed}[style=MyFrame,nobreak=true]
\begin{quote}
\small
``@\textless USER\textgreater He was allegedly slain by Soros, who then had Chavez's personal army of false voters cram him inside a Dominion voting machine before loading him into an RV with Hunter Biden's second laptop and Hillary's server.''

\end{quote}
\end{mdframed}
\descrit{5) Discussion:} it prompts discussion of the details of the misleading claim by adding commentary.
\begin{mdframed}[style=MyFrame,nobreak=true]
\begin{quote}
\small
``@\textless USER\textgreater What happened to all the votes cast for Trump that were destroyed? How are those tallied? Detroit-based Democratic Party activist a local: boasts On FB: I threw out every Trump ballot I saw while working for Wayne County, Michigan. They number in the tens of thousands, as did all of my coworkers.''

\end{quote}
\end{mdframed}
\descrit{6) Inquiry:} it inquires about the details of events related to the misleading claim and does not attempt to either support or deny the claim under question. 
\begin{mdframed}[style=MyFrame,nobreak=true]
\begin{quote}
\small
``Has anyone got a compelling justification for this? In accordance with a tweet I saw from @\textless USER\textgreater: A "James Bradley" born in 1900 has recently been entered into the Michigan Voter Information Center. James apparently submitted an absentee ballot on October 25. For a 120-year-old, not bad! <URL>''
\end{quote}
\end{mdframed}

\descrit{7) Irrelevant:} it is irrelevant to the misleading claim. These are considered false positives.
\begin{mdframed}[style=MyFrame,nobreak=true]
\begin{quote}
\small
``\#LOSANGELES: Our truck will be at the @HollywoodBowl voting location till 7pm. Use your voting rights and reward yourself with some.''
\end{quote}
\end{mdframed}

\begin{table}[t]
\centering
\small
	\setlength{\tabcolsep}{4pt}
\begin{tabular}{lrr}
\toprule
\textbf{Category} & \textbf{Candidates} & \textbf{Moderated (\%)} \\ \midrule
Amplifying        & 1,198               & 241 (20.11\%)                      \\
Reporting         & 922                & 222 (24.07\%)                      \\
Counter           & 122                & 4 (3.27\%)                         \\
Satire            & 15                 & 0 (0.00\%)                         \\
Discussion        & 646                & 83 (12.84\%)                       \\
Inquiry           & 84                 & 22 (26.19\%)                       \\
Irrelevant        & 59                 & 1 (1.69\%)                         \\ \bottomrule
\end{tabular}
\caption{Categories of candidate tweets and number/percentage receiving soft moderation by Twitter.}
\label{tab:candidate_category}
\end{table}

Table~\ref{tab:candidate_category}, reports the number of candidate tweets in each category. %
The vast majority falls in the amplification category with 1,198 out of 1,500 tweets (79.86\%), followed by tweets reporting about the false claim with 922 tweets (61.46\%).
43\% of the tweets add further discussion to the misleading claim under question rather than simply sharing the headline of a news article, and 8\% of them try to debunk it.
Finally, 59 (3.93\%) of the tweets flagged by \systemname are irrelevant to the claim under study and can therefore be considered false positives. 
As mentioned, the goal of \systemname is to flag tweets that are related to a claim that the platform wants to moderate, but human moderators should still make the final decision about applying labels to the candidates flagged by our system. 
We further discuss the implications of running \systemname in the wild in Section~\ref{sec:discussion}.

\begin{figure}
\centering
\includegraphics[width=0.95\columnwidth]{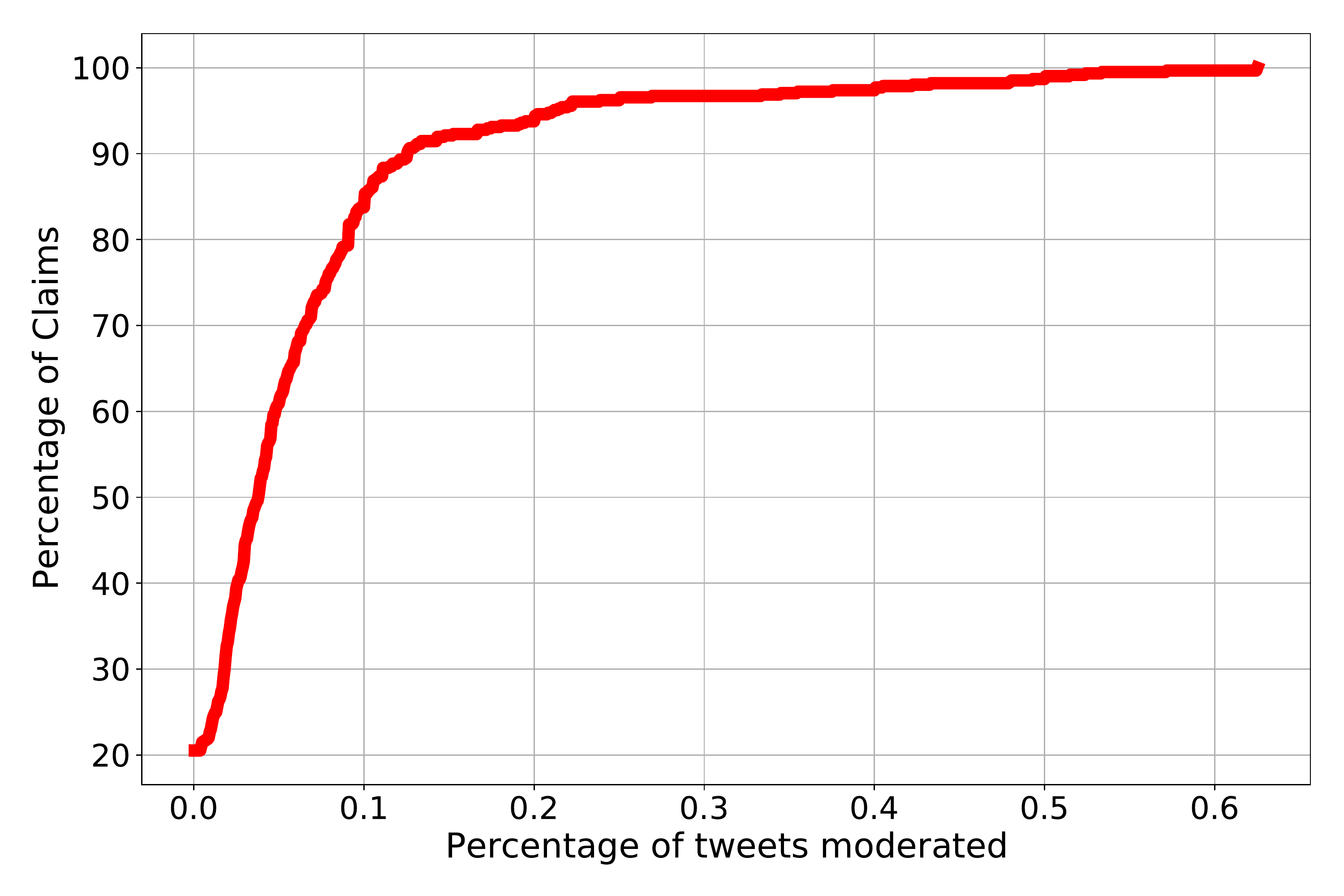}
\caption{Moderation coverage of misleading tweets flagged by \systemname per claim.}
\label{fig:coverage}
\end{figure}

\descr{False Negatives.} 
To evaluate the False Negatives of \systemname, we first evaluate the false negative of each of its two phases separately using \done~ for the first phase, and \dthree~ for the second phase.
In the claim structure extraction module, the Proposition Extractor component fails to extract 2.77\% of the propositions that are claim span.
After the propositions are extracted, \systemname misclassifies 3.46\% of the propositions that contain a claim, implying the missed claim structure would not be processed further in the second phase.
In the second phase, we quantify the proportion of tweets missed by the keywords identified through the LTR component of \systemname. 
The keywords produced by LTR identify 8,748 of the 10,776 tweets in the ground truth; this yields an 18.81\% false negative rate from \systemname's keyword extraction phase.
This is much lower than the false negative rate of the second best state-of-the-art approach, YAKE, which is 32.45\%.

\noindent{\bf Comparison to Twitter's soft moderation.}
After determining that the recommendations made by \systemname are accurate, we check if the tweets recommended by our approach were also soft-moderated by Twitter.
For every claim from the Claim Extraction Module, we retrieve the relevant set of tweets guided by the best set of keywords from our LTR component. 
We then follow~\cite{zannettou2021won} and extract metadata of soft moderation interventions for each tweet (i.e., if the tweet received a soft moderation and the corresponding warning label).
We perform this experiment on \dthree~.

Out of the 101,353 tweets flagged by \systemname as candidates for moderation, we find that only 4,330 (4.31\%) were soft moderated by Twitter.
Note that we could not check the existence of warning labels for 993 tweets as they were inaccessible, with either the tweets having been deleted or the accounts that posted them being deleted or suspended.
This experiment highlights the limitations of Twitter's soft moderation approach, suggesting that the platform would benefit from an automated system like \systemname to aid content moderation.
In Section~\ref{sec:twitter}, we further investigate whether we can identify a specific strategy followed by Twitter in moderating content.

\subsection{What drives Twitter moderation?}
\label{sec:twitter}

The analysis from the previous sections shows that Twitter only moderates a small fraction of tweets that should be moderated.
In this section, we aim to better understand how these moderation decisions are made.

We start by examining whether certain claims are moderated more aggressively than others and whether the type of message in a tweet affects its chances of being moderated. 
We then analyze the text and the URLs in moderated and unmoderated tweets, aiming to ascertain: 1) whether Twitter uses text similarity to identify moderation candidates and 2) whether Twitter automatically moderates all tweets linking to a known misleading news article.
Next, we look at the account characteristics of the users who posted moderated and unmoderated tweets, and engagement metrics (i.e., likes and retweets), aiming to understand if Twitter prioritizes moderating tweets by popular accounts or viral content.

\descr{Coverage by claim.}
In Figure~\ref{fig:coverage}, we plot the Cumulative Distribution Function (CDF) of the percentage of tweets moderated by Twitter for each of our 900 claims, out of the total candidate set flagged by \systemname.
Approximately 80\% of the claims have less than 10\% of the tweets moderated, whereas 95\% of claims have close to 20\% of the tweets moderated.
Very few claims (5) have at least half of the tweets moderated.
The misleading claim with the highest coverage is ``\textbf{\textit{Russ Ramsland file affidavit showing physical impossibility of election result in Michigan}}'' with 159 out of 309 (51\%) candidate tweets receiving moderation labels by Twitter.
On the other hand, the claim ``\textbf{\textit{Chinese Communists Used Computer Fraud and Mail Ballot Fraud to Interfere with Our National Election}}'' only has 1 out of 236 tweets (0.42\%) with warning labels.
This shows that, while the fraction of pertinent tweets moderated by Twitter is generally low, the platform seems to moderate certain claims more aggressively than others. 

\descr{Coverage by tweet type.}
In Section~\ref{sec:validation}, we list seven categories of tweets discussing misleading claims.
We now set out to understand whether Twitter moderates certain types of tweets more than others.
Table~\ref{tab:candidate_category} shows the fraction of tweets in our sample set of 1,500 manually analyzed tweets that did receive soft moderation by Twitter, broken down by category.
Tweets raising questions, reporting, or amplifying false claims are more likely to be moderated (with 26.19\%, 24.07\%, and 20.11\% of their tweets being moderated, respectively).
Satire tweets never received moderation labels, while tweets debunking false claims were only moderated in 3.27\% of the cases.
This indicates that Twitter considers the stance of a tweet mentioning a false claim, perhaps as part of a manual moderation effort.

\descr{Content analysis.}
Next, we investigate whether Twitter looks at near identical tweets when applying soft moderation decisions.
We take all tweets flagged as candidates by \systemname, and group together those with a high Jaccard similarity of their words.
We remove all the links, user mentions, and lemmatize the tweet tokens by using the \textbf{\textit{ekphrasis}} tokenizer~\cite{ekphrasis}.
We consider two tweets to be near identical if their Jaccard similarity is in the range 0.75--0.9 (out of 1.0).
We do so to extract tweet pairs that are not exactly the same, but have some variation in the content while discussing the same misleading claim.
We exclude retweets, and only consider the tweets originally authored by the users.

\begin{figure*}[t]
\centering
     \begin{subfigure}{0.375\linewidth}\includegraphics[width=\linewidth]{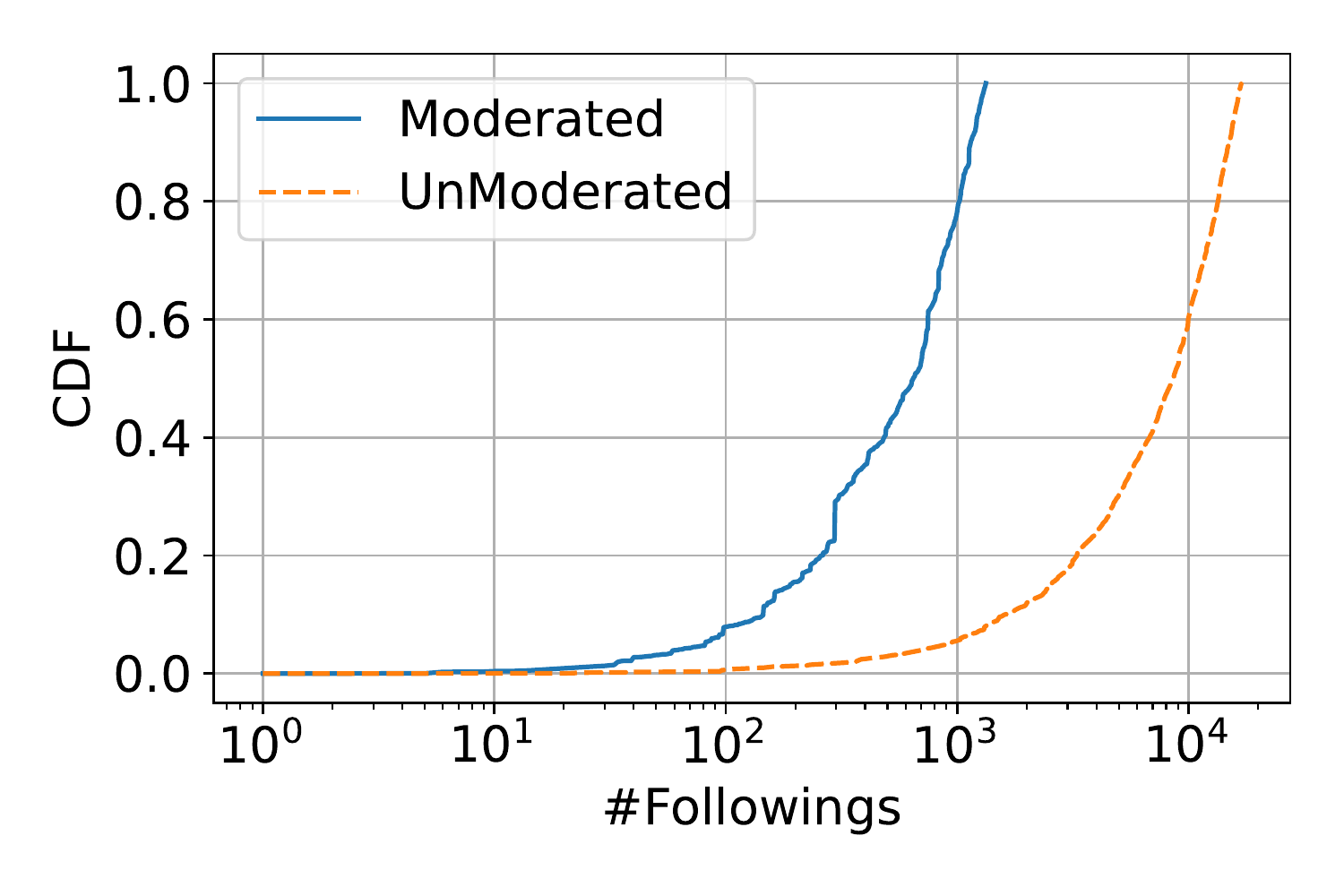}
     \caption{Following}          
	\end{subfigure}
	~
	 \begin{subfigure}{0.375\linewidth}\includegraphics[width=\linewidth]{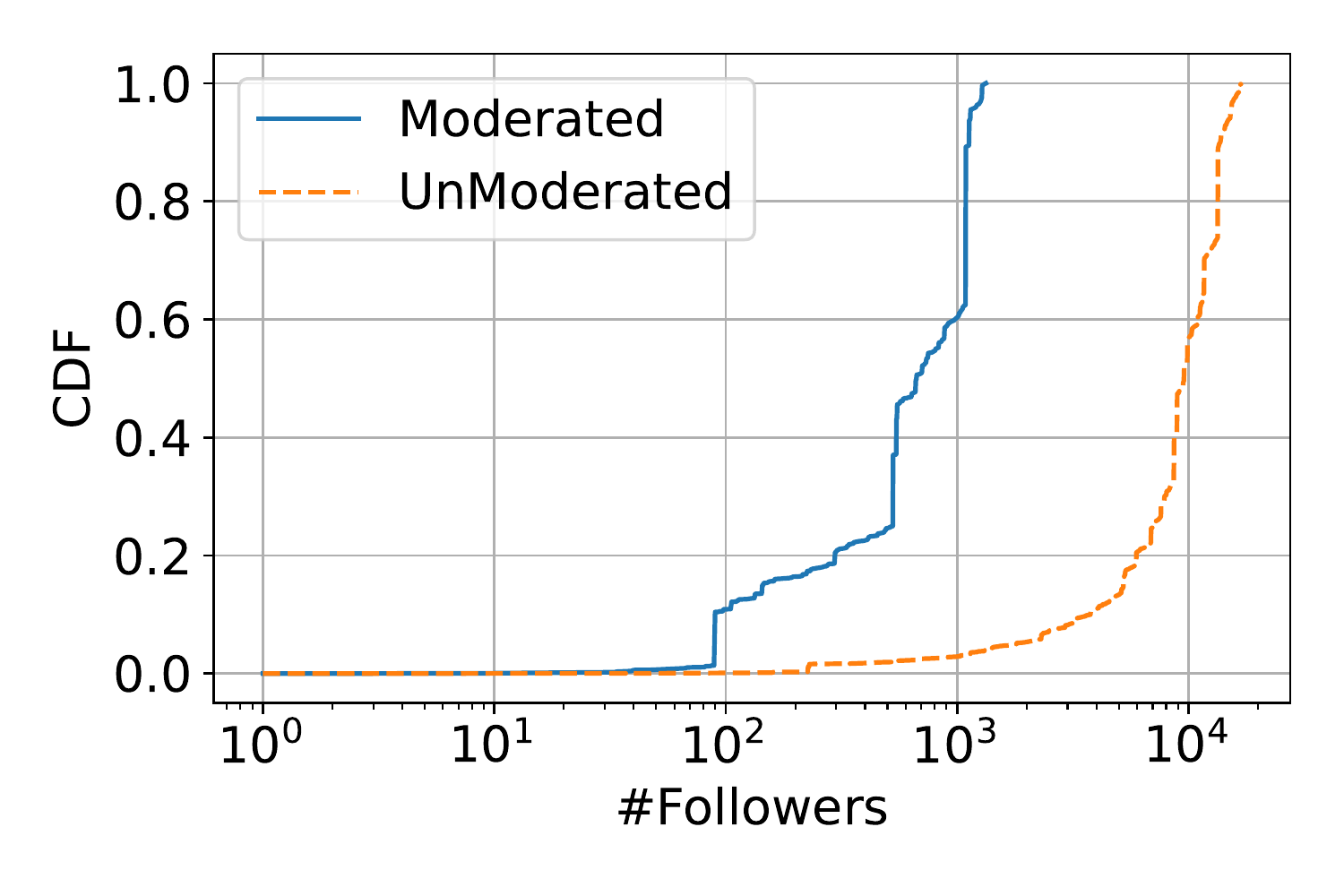}     
     \caption{Followers}      
	\end{subfigure}
	 \begin{subfigure}{0.375\linewidth}\includegraphics[width=\linewidth]{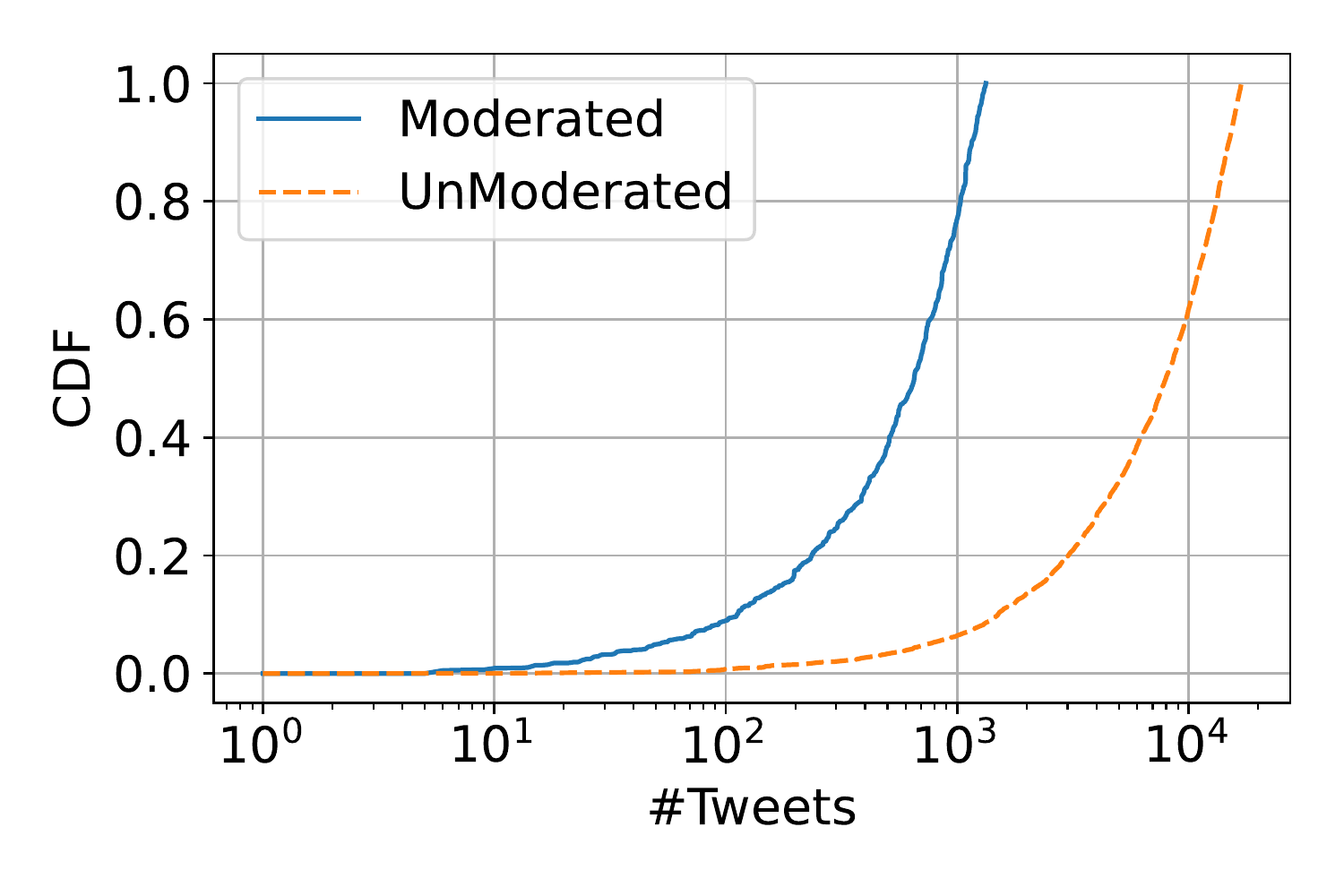}
     \caption{Tweet Counts}           
	\end{subfigure}
	~
	 \begin{subfigure}{0.375\linewidth}\includegraphics[width=\linewidth]{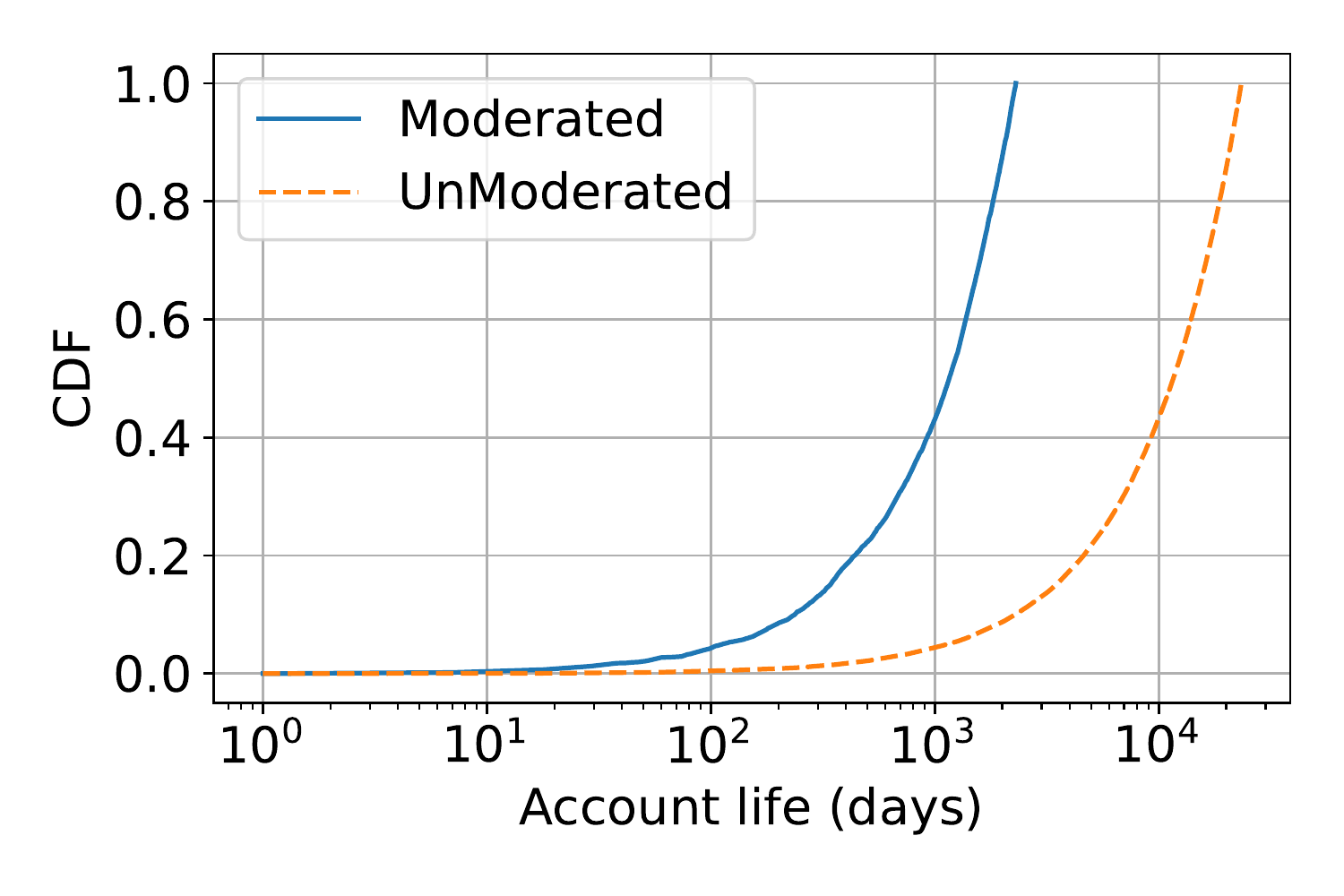}
     \caption{Account Age}           
	\end{subfigure}
	    
\caption{Cumulative Distribution Functions (CDF) of various user metrics for moderated and unmoderated tweets.}
\label{fig:ccdf_useranalysis}
\end{figure*}

We extract 17,241 pairs of tweets (out of 438,986 possible pairs), where at least one of the two was moderated by Twitter.
Only 3,857 pairs have both tweets moderated.
Note that \systemname effectively identifies {\em all} the 17,241 pairs of tweets as moderation candidates. %
Here is an example of a very similar pair of tweets, for which Twitter did not add labels to one of them:
\smallskip
\begin{mdframed}[style=MyFrame,nobreak=true]
\begin{quote}
\small
	\textbf{Moderated}:  ``RudyGiuliani in Trump campaign news conference: “"Joe Biden said a few weeks ago that his voting fraud crew was the best in the world. They were excellent, but we got them!”''
	
	\textbf{Unmoderated}: ``Joe Biden said a few weeks ago that his crew was the greatest in the world at catching voter fraud, but we caught them.''
\end{quote}
\end{mdframed}

These findings indicate that the decision by Twitter to add soft moderation to a tweet does not seem to be driven by the lexical similarity of tweets.

\descr{URL analysis.}
Another potential indicator used by Twitter when deciding which tweets to moderate is whether they include links to known news disinformation articles.
First, we expand all the links in the body of candidate tweets identified by \systemname to get rid of URL shorteners~\cite{maggi2013two}.
This yields 13,108 distinct URLs.
Next, we group candidate tweets by URLs and check what fraction of tweets sharing the URL are moderated by Twitter.

\begin{table}[t]
\centering
\small
	\setlength{\tabcolsep}{4pt}
\begin{tabular}{lrr}
\toprule
\textbf{URL news story}                  & \textbf{Candidates} & \textbf{Moderated} \\ \midrule
USPS whistleblower                       & 315                 & 7 (2.2\%)          \\
China manipulating election              & 252                 & 4 (1.5\%)          \\
Michigan ballot dump                     & 215                 & 44 (20\%)          \\
\#Suitcasegate related FB video & 208                 & 3 (1.4\%)          \\
Dominion remote machine control          & 135                 & 15 (11\%)          \\ \bottomrule
\end{tabular}%
\caption{Examples of URLs in candidate tweets and those being moderated by Twitter.}
\label{tab:urlmoderation}
\end{table}

Table~\ref{tab:urlmoderation} shows the five most common URLs (abstracted to the topic of the news articles) in our dataset, with the fraction of tweets including those URLs moderated by Twitter.
All these news stories, excluding one Facebook video, originate from known low-credibility websites like TheGatewayPundit and DC Dirty Laundry, which promote election misinformation.
Twitter moderates tweets containing those URLs in an inconsistent matter. %
Also note that \systemname can help identify 4,598 additional moderation candidate tweets compared to those on which Twitter intervened.

\descr{User analysis.}
We examine the differences in the social capital (e.g., number of followers) of the authors of tweets moderated by Twitter, compared to those our system recommends for moderation but for which Twitter did not intervene.
Figure~\ref{fig:ccdf_useranalysis} reports the CDF of followers, following, tweet count, and account age of accounts that posted moderated and unmoderated tweets.
We find that authors of tweets that have warning labels have much fewer followers, followings, lower account activity, and have younger accounts than tweets without warning labels.
We also conduct two-sample Kolmogorov-Smirnov tests for each user metric, finding that the differences are statistically significant for followers and account age ($p < 0.01$) as well as following count and status count ($p < 0.05$).
This goes against the notion that popular accounts are more likely to have their content moderated.

We also check if the accounts with moderated tweets were suspended for violating Twitter Rules~\cite{twitter_rules}. %
We find that only 33 out of 3,397 users were suspended by Twitter;  this gives us strong ground to rule out the possibility that tweet moderation is not due to the ``legitimacy'' of the account themselves.

\begin{figure}
\centering
     \begin{subfigure}{0.375\textwidth}\includegraphics[width=\linewidth]{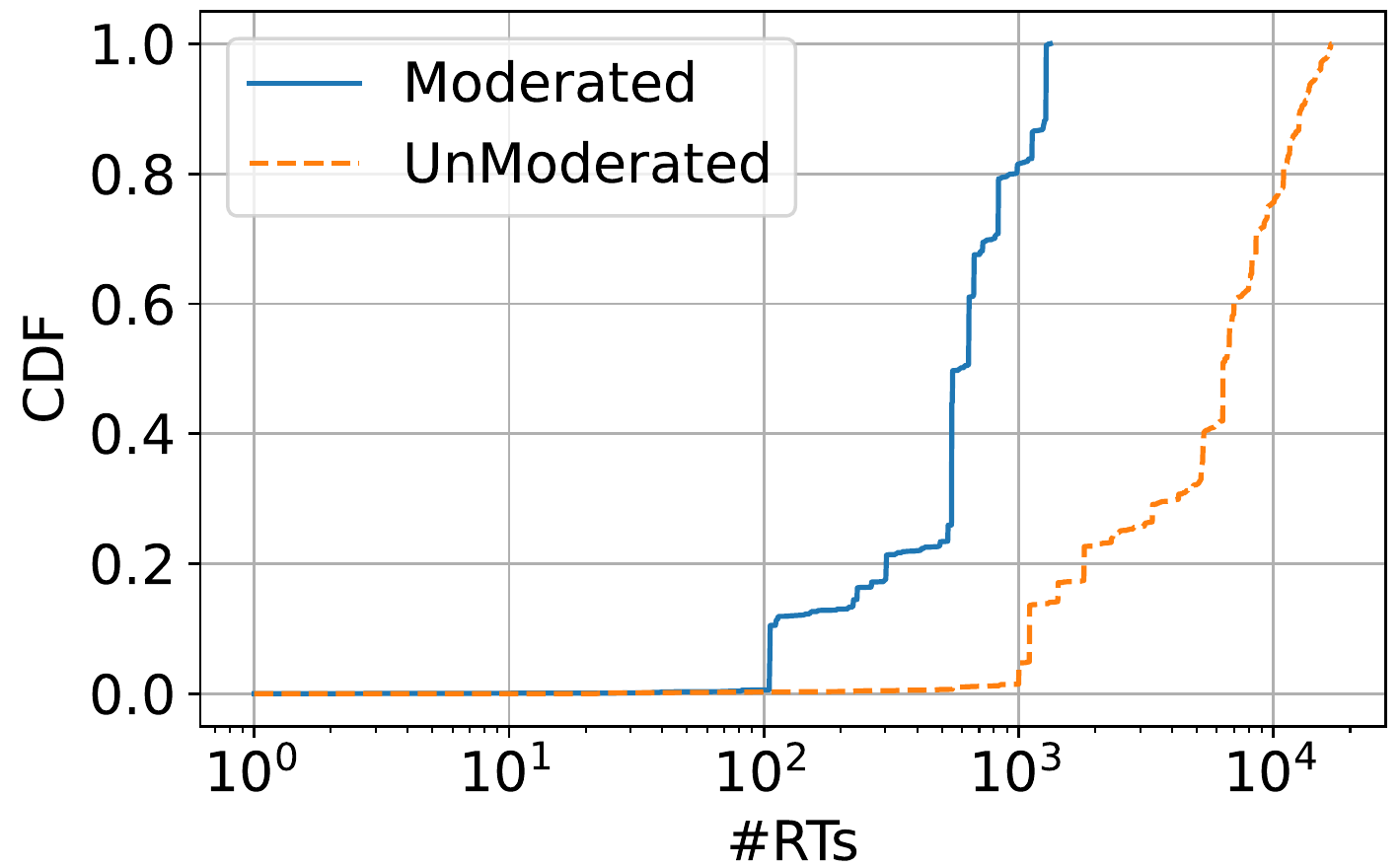}     
     \caption{Retweets}     
	\end{subfigure}
	 \begin{subfigure}{0.375\textwidth}\includegraphics[width=\linewidth]{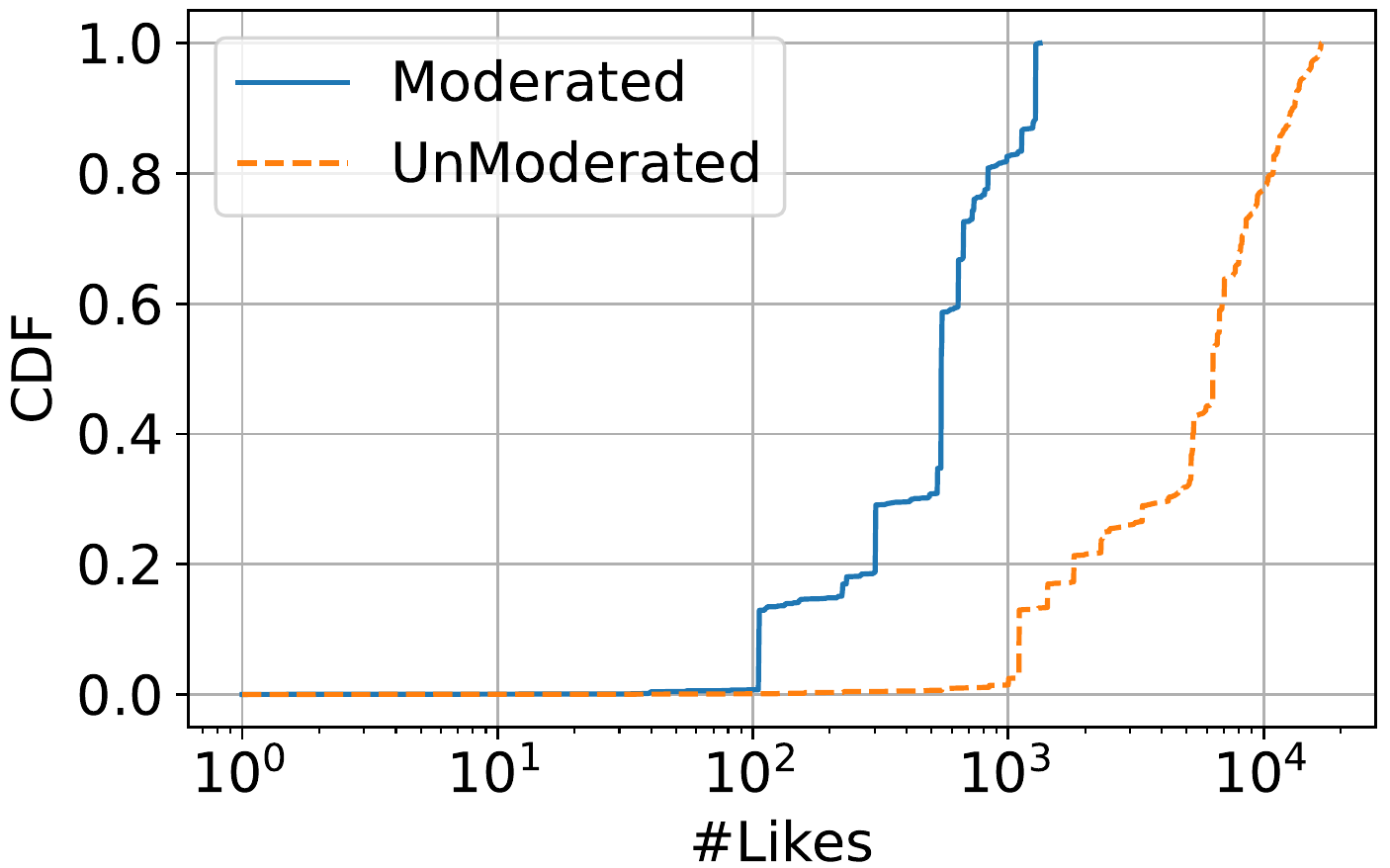}  
     \caption{Likes}         
	\end{subfigure}
	    
\caption{CDFs of engagement metrics for moderated and unmoderated tweets.}
\label{fig:ccdf_tweetanalysis}
\end{figure}

\descr{Engagement analysis.}
Finally, we analyze engagement metrics. %
Figure~\ref{fig:ccdf_tweetanalysis} reports the CDF of retweets and likes categorized by moderation status of the 101,353 candidate tweets \systemname flags for moderation from \dthree~, compared to the ones flagged by Twitter.
Similar to the user analysis, we find that unmoderated tweets have more engagement. %
When we check for statistical significance of difference in distributions of the retweet count using Kolmogorov-Smirnov tests, we find that it is statistically significant ($p < 0.01$), while we cannot reject the null hypothesis for the likes.
Note, however, that these results have to be taken with a grain of salt, as we do not have a timeline of when exactly moderation was applied, and whether the soft interventions hampered the virality of online content. 

\descr{Takeaways.} Our analysis paints a puzzling picture of soft moderation on Twitter. 
We find that certain claims are moderated more aggressively.
Still, Twitter does not seem to have a system in place to identify similar tweets discussing the same false narrative, nor flagging tweets that link to the same debunked news article.
We also find that Twitter does not appear to focus on the tweets posted by popular accounts for moderation, but rather that tweets posted by accounts with more followers, friends, activity, and a longer lifespan are more likely to go unmoderated.
This confirms the need for a system like \systemname. %

\section{Related Work}
In this section, we review relevant work on soft moderation, security warnings, and keyword extraction in the context of disinformation.

\descr{Soft Moderation during the 2020 Elections.}
As part of the Civic Integrity Policy efforts surrounding the 2020 US elections, Twitter applied warning labels on ``misleading information.''
Empirical analysis~\cite{zannettou2021won} reports 12 different types of warning messages occurring on a sample of 2,244 tweets with warning labels.
Statistical assessment of the impact of Twitter labels on Donald Trump's false claims during the 2020 US Presidential election finds that warning labels did not result in any statistically significant increase or decrease in the spread of misinformation~\cite{papakyriakopoulos2022impact}.
Twitter later reported that approximately 74\% of the tweet viewership happened post-moderation and, more importantly, that the warnings yielded an estimated 29\% decrease in users quoting the labeled tweets~\cite{twitter_update_2020}. %

\descr{Security Warnings for Disinformation.}
The warning labels adopted by Twitter as soft moderation intervention can be broadly categorized as a type of security warning.
Security warnings can be classified into two types: contextual and interstitial.
The former passively inform the users about misinformation through UI elements that appear alongside social media posts.
The latter prompt the user to engage before taking action with the potential piece of disinformation (e.g., retweeting or sharing).
A recent study~\cite{kaiser2021adapting} shows that interstitial warnings may be more effective, with a lower clickthrough rate of misleading articles.
Additionally, interstitial warnings are more effective design-wise because they capture attention and provide a window of opportunity for users to think about their actions.
Efforts to study warning labels on countering disinformation have thus far been mostly focused on Facebook~\cite{pennycook2018prior,ross2018fake,pennycook2020implied}, where warning labels were limited to ``disputed'' or ``rated false'', and the approach was deemed to be of limited utility by Facebook~\cite{smith2017designing}.
Recently, other platforms like Twitter~\cite{alba2020twitter}, Google~\cite{googlelabel}, and Bing~\cite{bing_label} also used some form of fact-check warnings to counter disinformation.

\descr{Tools for automated content moderation.}
The sheer scale of content being produced on modern social media platforms (Facebook, Reddit, YouTube etc.) have motivated the need to adopt tools for automated content moderation~\cite{fb_guidelines,youtube_guidelines}.
However, due to the nuanced and context-sensitive nature of content moderation, it is a complex socio-technical problem~\cite{jhaver2019human,seering2019moderator}.
Most of the work in this space of automated content moderation are focused on Reddit, aiming to identify submissions that violate community-specific policies and norms ranging from hate speech to other types of problematic content~\cite{seering2020reconsidering}.
The most popular solution to automated content moderation in Reddit, AutoModerator~\cite{jhaver2019human} allows community moderators to set up simple rules based on regular expressions and metadata of users for automated moderation.  
On YouTube, FilterBuddy~\cite{jhaver2022designing} is available as a tool for creator-led content moderation by designing filters for keywords and key phrases.
Similarly, Twitch offers an automated moderation tool called Automod to allow creators to moderate four categories of content (discriminations and slurs, sexual content, hostility, and profanity) on the platform~\cite{twitch_automod}.
Another tool, called CrossMod~\cite{chandrasekharan2019crossmod} uses an ensemble of models learned via cross-community learning from empirical moderation decisions made on two subreddits of over 10M subscribers each.

\descr{Keyword Extraction for Disinformation.}
Researchers have used an array of methods to detect disinformation, ranging from modeling user interactions~\cite{shu2019role,tschiatschek2018fake,qian2018neural}, leveraging semantic content~\cite{ciampaglia2015computational,zhang2017constraint,esmaeilzadeh2019neural,pan2018content}, and graph based representations~\cite{nguyen2020fang,gangireddy2020unsupervised,lu2020gcan}.
The foundation of our system lies in the keyword detection, which has been used before to study disinformation on social media.
DisInfoNet, a toolbox presented in~\cite{guarino2019beyond}, represents news stories through keyword-based queries to track news stories and reconstruct the prevalence of disinformation over time and space.
Similarly, the work in~\cite{gaglani2020unsupervised} uses keyword extraction techniques as the base for semantic search to detect fake news on WhatsApp.
The work in~\cite{choudhary2018neural} focuses on credibility assessment of textual claims on news articles with potentially false information, also using keyword extraction as a part of their multi-component module.

\descr{Learning To Rank for Keyword extraction.}
The closest applications of LTR to our work are the proposals in~\citep{jiang2009ranking} for keyphrase extraction and in~\citep{cai2017keyword} for keyword extraction in Chinese news articles.
The foundational work by~\citep{jiang2009ranking} motivates the necessity of framing the problem of keyphrase extraction as a ranking task rather than a classification task while improving results on extracting keyphrases from academic research papers, and social tagging data.
The LTR approach for keyword extraction utilized by \systemname is motivated by the premise set up by this work.
Similarly,~\citep{cai2017keyword} use Learning To Rank to identify keywords from 1800 public Chinese new articles using TF-IDF, TextRank, and Latent Dirichlet Allocation (LDA) as the set of features for the ranking model.

\section{Discussion and Conclusion}
\label{sec:discussion}

This paper presented \systemname, a system geared to automatically flag candidate tweets for soft moderation interventions.
Our experiments demonstrate that Learning to Rank (LTR) techniques are effective, that \systemname outperforms other approaches, produces accurate recommendations, and can increase the set of tweets that receive soft moderation by over 20 times compared to those flagged by Twitter during the 2020 US Presidential Election.

\descr{Implications for social media platforms.}
As discussed in Section~\ref{sec:twitter}, soft moderation interventions applied by Twitter appear to be spotty and not following precise criteria.
This might be due to moderation being conducted mainly in an ad-hoc fashion, relying on user reports and the judgment of moderators.
\systemname can assist this human effort, working upstream of the content moderation process and presenting moderators with an optimal set of tweets that are candidates for moderation. 
Because of the nuances of moderating false and misleading content, we envision \systemname to be deployed as an aid to human moderation rather than an automated detection tool.

Nonetheless, the claim-specific design of \systemname can also be used by moderators for other actions as per their policies, e.g., asking users to remove a given tweet or performing hard moderation by removing the tweets.
The choice between soft moderation and hard moderation can be made by moderators contextually after the moderation candidates are retrieved through \systemname, either based on underlying claims or a case-by-case basis.
E.g., platforms may decide to soft moderate posts that push a certain false narrative but not add warnings if posts inform users about falsehood.
Alternatively, they might add warnings to posts about the false narrative, providing additional context to users and allowing them to make up their minds about it.
Platforms could also craft warning messages depending on the context in which a false claim is discussed or design these messages to be more effective based on the audience and risk levels of specific false claims.
For example, different type of warning messages can be applied by platforms to distinguish between different levels of risk associated with the misleading claims (e.g., high and low-level risks associated with COVID-19 misinformation)~\cite{ling2022learn}.
We are confident that Human-Computer Interaction researchers will be able to address these challenges, which go beyond the scope of this paper.

\descr{Human effort required for adopting \systemname.}
When setting up \systemname to work in a new context, platform moderators need to follow the steps highlighted in Sections~\ref{sec:claimextraction} and~\ref{sec:keyword}.
First, they need a set of tweets with claims they identified as containing misleading information, together with a tuning dataset like \dthree~.
Moderators can create a tuning dataset like \dthree~ by using a broad set of keywords associated with the event or topic and querying the Twitter API to which they have full access (e.g. ``COVID-19,''  ``coronavirus,'' etc., in the case of the pandemic).
They then need to tune the threshold for the Claim Stopper API in the Claim Extraction component (see Section~\ref{sec:claimextraction}).
In our experiments, this phase took us, on average, two minutes per claim.
Finally, they need to create the training set for the LTR model by following the iterative process discussed in Section~\ref{sec:keyword}.
When performed by a single annotator, this process took, on average, 15 minutes per claim for the experiments discussed in this paper.
Twitter could speed up these steps further by having multiple annotators work on the same task.
Additionally, the work required on each claim is independent of other claims; therefore this process can be easily parallelized within the organization or even through crowdsourcing campaigns~\cite{founta2018large,lease2012crowdsourcing,oleson2011programmatic}.

\descr{Resilience to evasion.}
As with any adversarial problem, malicious actors are likely to try to evade being flagged by \systemname.
E.g., they might avoid using certain words to avoid detection and use synonyms or dog whistles instead~\cite{gerrard2018beyond,tahmasbi2021go,zannettou2020quantitative,zhu2021self}.
However, this would make the false messaging less accessible to the general public, who would need to first understand the alternative words used and ultimately be counterproductive for malicious actors by limiting the reach of false narratives.

\descr{Limitations.} \systemname requires a seed of tweets to be moderated, making it inherently reactive.
However, this is a problem common to all moderation approaches, including the work conducted by fact-checking organizations.
Another limitation is that we could only test \systemname on one dataset related to the same major event (the 2020 US Presidential Election), as this is the only reliable dataset with soft moderation labels available to the research community.

Even though Twitter applied warning labels on misinformation about COVID-19, previous research reported that these were unreliable and inconsistent~\cite{lange_2020,lyons_2020}, which we independently confirmed in our preliminary analysis.
More recently, Twitter recently started applying warning labels to tweets in the context of the Russian invasion of Ukraine~\cite{benson_twitter_russia}, but these labels are applied based on the account posting them (i.e., if the account belongs to Russian or Belarusian state-affiliated media) instead of being claim-specific as required by \systemname.
While the LTR model used by \systemname is not specific to the actual keywords being searched, and therefore we expect that it should generalize across the entirety of Twitter, platform moderators using the tool should take further steps to validate it when used in contexts other than politics and elections.

\descr{Future work.} 
We plan to extend \systemname to additional platforms.
Since our system only needs the text of posts as input, we expect it to generalize to other platforms, e.g., Facebook, Reddit, etc.
We will also investigate how claims automatically built by \systemname can be incorporated into warning messages to provide more context to users and allow them to be better protected against disinformation.

\descr{Acknowledgments.} 
We thank the anonymous reviewers for their comments that helped us improve the paper.
Our work was supported by the NSF under grants CNS-1942610, IIS-2046590, CNS-2114407, IIP-1827700, and CNS-2114411, and by the UK's National Research Centre on Privacy, Harm Reduction, and Adversarial Influence Online (REPHRAIN, UKRI grant: EP/V011189/1).

\small
\bibliographystyle{abbrv}
%\bibliography{refs}

\end{document}